\documentclass[aps,pra,twocolumn,a4paper,nofootinbib,preprintnumbers] {revtex4-1}
\usepackage{graphicx, color}
\usepackage[dvipsnames]{xcolor}
\usepackage[colorlinks=true,urlcolor=blue,citecolor=blue,linkcolor=blue]{hyperref}
\usepackage{amsmath}
\usepackage{amssymb}
\usepackage{amsbsy}

\newcommand{\D}{{\rm{d}}}
\newcommand{\I}{{\rm{I}}}

\begin{document}

\preprint{PHYSICAL REVIEW A \textbf{99}, 053830  (2019)}

\title{Satellite-mediated quantum atmospheric links}

 \author{D. Vasylyev}
 \affiliation{Institut f\"ur Physik, Universit\"at Rostock,
 Albert-Einstein-Stra\ss{}e 23, 18059 Rostock, Germany}

 \author{W. Vogel}
 \affiliation{Institut f\"ur Physik, Universit\"at Rostock,
  Albert-Einstein-Stra\ss{}e 23, 18059 Rostock, Germany}

   \author{F.~Moll}
 \affiliation{Institute of Communications and Navigation, German Aerospace Center, 82234 Wessling, Germany}

\begin{abstract}
The establishment of quantum communication links over a global scale is enabled by  satellite nodes.
We examine the influence of Earth's atmosphere on the performance of quantum optical communication channels with emphasis on the downlink scenario.
We derive the geometrical path length between a moving low Earth orbit satellite and an optical ground station as a function of the ground observer's zenith angle, his geographical latitude, and the meridian inclination angle of the  satellite.
We show that the signal distortions due to regular atmospheric refraction, atmospheric absorption, and turbulence   have a strong dependence on the zenith angle.
The observed saturation of transmittance fluctuations for large zenith angles is explained.
The probability distribution of the transmittance for slant propagation paths is derived, which enables us to perform the security analysis of decoy state protocols implemented via satellite-mediated links.
\end{abstract}
\pacs{}

\maketitle


\section{Introduction}

The recent success in practical realization of quantum state transfer between satellites and ground stations~\cite{Vallone2015, Vallone2016, Yin2017, Yin2017a, Guenthner2017, Liao2017, Liao,  Takenaka, Liao2018, Bedington2017, Calderaro2018} has established the cornerstone of future global quantum communication.
This achievement is especially impressive in the view of how fast the quantum communication over  free-space quantum links reached its maturity.
During only three decades the free-space quantum communication reached distances of 7600 km~\cite{Liao2018}, while  starting from proof-of-principle demonstration over a 30-cm quantum channel~\cite{Bennett1989}.

An acceleration of the development of satellite-based quantum technologies is caused by the need to establish global secure communication networks.
The present classical public key cryptography is based on mathematical problems that admit no efficient solutions with currently available technologies.
However, its security is vulnerable to quantum computer hacking attacks, in particular, those based on Shor's factoring algorithm~\cite{Shor1994}.
Therefore, the future  realization of quantum computers threatens the currently used classical cryptographic protocols  to become insecure and hence useless.

Conversely, quantum key distribution (QKD) establishes unconditional secure cryptographic keys between two distinct parties~\cite{Bennett1984, Scarani2009}.
The security of QKD is based on some fundamental principles of quantum physics such as the no-cloning theorem \cite{Wooters1982}, Bell correlations~\cite{Ekert1991}, uncertainty relations for the most of continuous-variable protocols \cite{Hillery, Cerf, Gottesman, Grosshans2002, Grosshans2003, Madsen, Braunstein}, etc.
Ironically, while providing the security of QKD, the no-cloning theorem at the same time puts limitations on possible communication distances due to inability to amplify a quantum signal.
One way to bring quantum communication to global scale is the use of satellites.
Satellite-mediated QKD networks could span large communication distances  by linking widely separated ground stations~\cite{Liao2018, Bedington2017}.

The setup of the satellite-mediated quantum links is a demanding task.
The following problems must be overcome for successfully establishing a feasible quantum communication with satellites: the relative motion of the communication parties \cite{Bonato2006, Zhang2014, Pramanik2017}, the influence of gravity~\cite{Bruschi2014, Bruschi2018}, the clock synchronization problem \cite{Robert2016, Wang2016}, acquisition, tracking and  pointing issues with moving platforms~\cite{Nauerth2013, Wang2013, Pugh2017}, the influence of background noise \cite{Erlong2005, Bonato2009, Tomaello2011}, to name just a few.
For low Earth orbiting (LEO) satellites, the communication time is limited to a few minutes~\cite{Bacco2013, Casado2017} and this puts additional tight bounds on communication security.
Finally, Earth's atmosphere contributes significantly to the loss budget and to the deterioration of the quantum signal due to diffraction, scattering, absorption, and atmospheric turbulence~\cite{Tatarskii2016, Andrews2005, Bourgoin2013}.

The common feature of the aforementioned issues is the strong dependence of disturbing factors on the relative position of the moving satellite and the ground station.
For the observer at the ground station the instantaneous position of the satellite is given by the zenith angle, i.e., by the angle between the satellite and the vertical direction.
The dependence of atmospheric disturbances on the zenith angle has attracted the attention of scientists since ancient times.
Already Ptolemy (100-175 A.D.)  noticed that the atmospheric refraction displaces the apparent position of celestial bodies toward the zenith~\cite{Lehn2005}.
Such a regular refraction is caused by the altitude variation of the atmospheric refractive index.
It influences the precision of satellite measurements and is responsible for apparent elongation of celestial objects, mirages, green flashes on setting sun, to name just a few~\cite{Young2006}.
Other types of  disturbances, such as intensity scintillation~\cite{Andrews2000, Dravins1997} and image dancing~\cite{Tatarskii2016, Chiba1971}, arise from turbulent fluctuations of the atmospheric refractive index.
The strength of these fluctuations varies with altitude and hence its influence on optical light propagation is also a function of the zenith angle.
Thus, the examination of the zenith angle dependence of relevant characteristics is of immense importance for the establishing of robust and reliable optical quantum communication links via the satellites.

In this article we address several aspects of ground-to-space communication  connected with Earth's atmosphere.
We focus  on quantum links between  optical ground stations (OGS) and LEO satellites.
Orbital periods of LEO satellites are of 120 min or less, leading to short passage times of the OGS, with which communication is aimed to be established.
Depending on the inclination angle of the satellite orbit, the expected communication time varies in the range of 8-16 min.
This communication time window is actually narrower due to the time needed for satellite acquisition.
Additionally, satellites with trajectories of small  inclination angle could lose the contact with the OGS because of natural barriers, while  OGS telescopes could disconnect communication with  satellites at small zenith angles due to  mechanical construction of telescope mounts.
Under these circumstances, it is desirable to start the acquisition procedure when the satellite appears on the horizon.
In this case, quantum communication could be established already at small satellite elevation angles  (large zenith angles).
Since the disturbing influence of the atmosphere grows with the growth of the zenith angle, the zenith angle dependence is crucial for the strict analysis of quantum satellite communication links.

In the following we provide a systematic analysis of quantum communication channels with the inclusion of disturbance effects due to geometrical elongation of communication links, atmospheric regular refraction,  extinction, and turbulence.
In our consideration we focus on the zenith angle dependence of the associated signal losses.
Additionally, the finite communication time puts severe restrictions on   secure quantum key length in satellite-mediated communication.
Careful estimation of the loss budget of quantum channels is needed to estimate the lower bound of the key length and its rate.
In this article we perform the security analysis of a decoy state protocol that is commonly used in practice.

This article is structured as follows.
In Sec.~\ref{sec:Orbit} we review some basic communication scenarios with satellites and in Sec.~\ref{sec:distance} we obtain the geometrical distance between a ground station and a satellite.
We focus on atmospheric refraction effects in Sec.~\ref{sec:Refraction}.
The regular extinction of the optical signal due to scattering and absorption on air particles is discussed in Sec.~\ref{sec:DeterministicExt}.
In Sec.~\ref{sec:Turbulence}, the dependence of optical field correlation functions on the zenith angle for the optical field propagating in turbulence is studied.
These functions allow us to calculate the scintillation index, the mean beam-spot radius, and the beam wandering variance.
In Sec.~\ref{sec:PDT}, we derive the probability distribution of the transmittance that characterizes satellite-mediated atmospheric quantum channels.
An application of the developed theory to the security analysis of decoy-state protocols is discussed in Sec.~\ref{sec:Decoy}.
A summary and conclusions are given in Sec.~\ref{sec:Conclusions}.

\section{Satellite-mediated communication links}
\label{sec:Orbit}

A low Earth orbit (LEO) is an orbit around Earth with an altitude above Earth's surface  $H$ ranging from 160 to 2000 km and an orbital period between about 80 and 130 min.
The fast movement of the satellite puts certain restrictions on the communication performance with the ground stations and limits the communication window to several minutes.
The duration of the communication session depends  on OGS' geographical position, the duration of acquisition procedure, the presence of natural  obstacles along the optical path, etc.
The signal-to-noise ratio for satellite-mediated optical link is connected with the length of the optical path between the communication parties and hence it is related with the shape  of the satellite orbit and its relative position to  the ground-station communication party.
Moreover, atmospheric refraction bends the optical ray paths, while the atmospheric turbulence corresponds to various random refraction and diffraction phenomena.
In this section we remind the reader of some basic geometrical concepts of satellite-mediated optical links and leave the detailed analysis of some aforementioned aspects to the following sections.

\begin{figure}[ht]
 \includegraphics[width=0.48\textwidth]{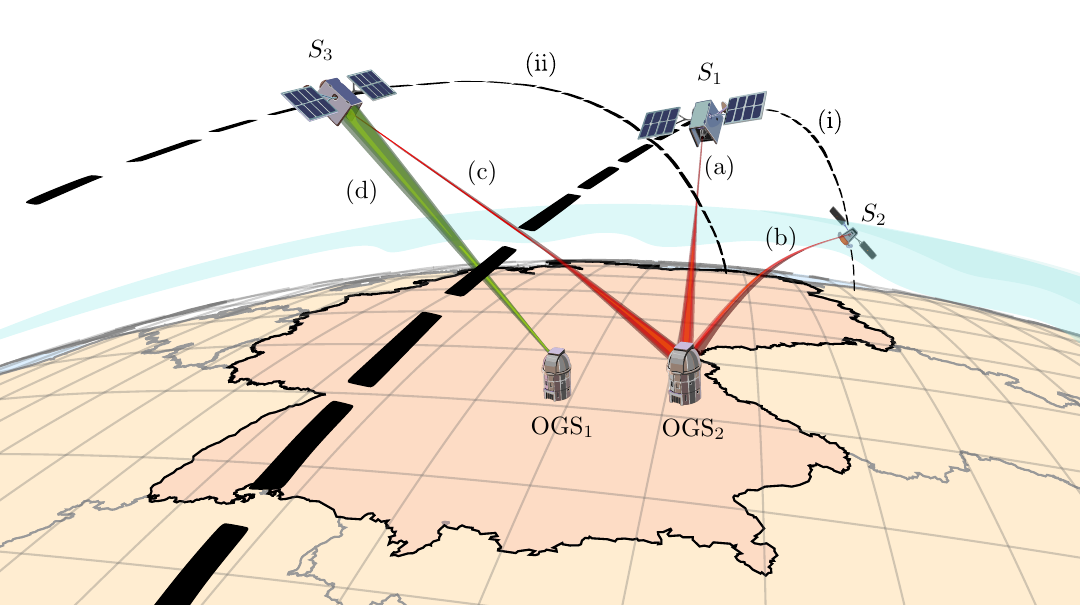}
 \caption{\label{fig:Satellites} Typical communication scenarios of optical ground stations (OGS) with satellites $S$ via uplink (d) and downlinks (a)-(c).
 The subscript $i=1,2,3$ in the notation $S_i$ refers to the successive positions of the satellite.
After the satellite  makes one orbital revolution along its orbit (i) its new position $S_3$ as well as orbit (ii) appear  inclined for the  ground observer OGS${}_2$.
This apparent change in the orbit inclination is due to the Earth rotation.
}
\end{figure}

Figure~\ref{fig:Satellites} shows the generic communication experiment with classical or quantum light.
Two ground stations OGS${}_1$ and OGS${}_2$ establish the communication links with satellite $S$ shown in successive moments of time.
The instantaneous positions of the satellite at times $t_1$, $t_2$, and $t_3$ are denoted as $S_1$, $S_2$, and $S_3$, respectively.
The satellite has a polar orbit which passes over the Earth's polar regions from  south to north.
Moreover, since orbits  of most  LEO satellites have vanishingly small ellipticity, we will consider  circular orbits in the following.
Initially at time $t_1$,  the orbit, denoted as (i), passes through the meridian plane of OGS${}_2$ but has some inclination to meridian plane of OGS${}_1$.
At later time $t_2$, the satellite approaches the horizon,  makes one full orbit revolution and appears at the point $S_3$.
For a ground observer\footnote{In the following we use the word "observer" in general referring to the communication party located in the optical ground station.}  the apparent satellite trajectory changes from (i) to (ii) during this single revolution period due to the Earth rotation.
In the following we will also assume that the observer's parallax due to Earth's rotation during one communication session can be neglected.
This is justified if the flight time over the observer horizon is much smaller than the satellite orbiting period\footnote{Strictly speaking, for satellites with small altitudes above the ground and for observers located at small geographical latitudes the corrections due to the aforementioned parallax effects should be included.}.

The satellite zenith trajectory that passes through the observer's meridian plane can be considered as the most favorable for establishing a communication link and is called as the "best pass" by some authors~\cite{Bourgoin2013}.
Indeed, in this case the  link length is smaller compared to the link's lengths for inclined satellites.
However, in many practical cases the mechanical mount of the OGS receiver or sender telescopes prevents one to use the whole advantage of the link that can lead to the interruption of the communication process near small zenith angles of the observer~\cite{Casado2017}.
In such cases the trade-off exists between the use of optimal propagation path and the duration of communication session.
In this context the analysis of communication with the satellites moving along the inclined orbits is important.

Depending on the location of the source we distinguish downlink [paths (a)-(c) in Fig.~\ref{fig:Satellites}] and uplink [path (d)] communication scenarios.
A downlink geometry is favorable for optical signal transmission since the optical beam starts to propagate in vacuum until it enters the atmosphere.
As a consequence the transmitted beam shows smaller diffraction-induced broadening and small beam wandering due to refraction on turbulent atmospheric inhomogeneities.
This is the reason why the majority of  quantum optical experiments are being performed in downlink configuration~\cite{Guenthner2017, Takenaka, Casado2017, Liao2018}.
Under comparable atmospheric conditions an uplink communication shows inferior performance due to the influence of atmospheric turbulence already on early stages of optical signal transmission.
Theoretical studies \cite{Bourgoin2013} reported the reduction in received key bits being less than one order of magnitude compared to downlink scenario.
However, an uplink for quantum communication purposes poses the following advantages: simple design of satellite missions or use of already launched satellites~\cite{Vallone2015},
relaxed requirements on data storage and processing equipment, variability of quantum light sources and their accessibility for maintenance and repair~\cite{Pugh2017}.
In this article, we will focus primarily on downlink communication scenarios.


\section{Regular losses of satellite-mediated link}
An optical Gaussian beam undergoes diffraction-induced broadening while propagating in free space.
The amount of broadening depends on propagation path length and whether beam is focused, collimated, or divergent.
Since the receiver collects the incoming light with the aid of finite-aperture device such as telescope, only a fraction of the signal reaches the detector.
For fixed aperture size, the signal losses due to  truncation on the aperture will increase with the increase of the propagation path.
Moreover, absorption and scattering on atmospheric gases and aerosols leads to the degradation of the signal-to-noise ratio as well.

In this section we consider the effects leading to the regular diffraction-induced and extinction losses in satellite-mediated links.
By the word "regular" we refer to effects that occur in a systematical manner.
Among such effects we distinguish purely geometric optical path elongation due to increase of relative positions of communication parties.
The regular losses in this case depend on the geographical location of the OGS, on the type of the satellite orbit, and on such characteristics as the satellite altitude, orbit inclination angle, and the satellite declination angle to the Equator plane.
Additionally, regular atmospheric refraction bends the light rays, increasing the optical propagation path and contributing to loss budget.
This effect is especially pronounced if the satellite is positioned close to the observer horizon.
Finally, we discuss the signal loss due to atmospheric absorption and scattering.


\subsection{Slant range }
\label{sec:distance}

We consider the communication scenario with an orbiting satellite which has been described in the previous section and derive the length of the line segment connecting the ground observer and the satellite.
This  purely geometric length referred as the slant range does not account for any elongation effects due to atmosphere  and depends on the geographical location of the observer and on the parameters of the satellite orbit.

\begin{figure}[ht]
 \includegraphics[width=0.45\textwidth]{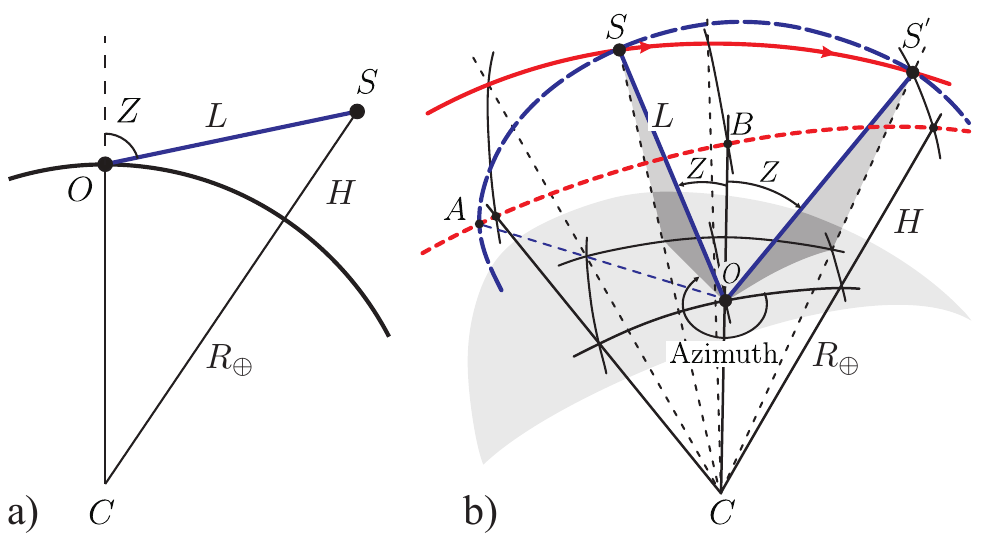}
 \caption{\label{fig:Geometry1}
Geometry of the communication configuration shown in the plane that passes through the Earth center $C$, the observer's location $O$, and the satellite position $S$ (a).
The same communication link $OS$ is shown in its relative position to the observer's meridian plane $ABOC$ (b).
The cross-section of the trajectory with the cone of angle $2Z$ and side  $SO$ yields the associated link $OS^\prime$ of the same length.
}
\end{figure}

In the following we consider a perfect polar satellite orbit that passes through both north and south poles.
For simplicity, we assume that the orbit is perfectly circular with the radius $R=R_\oplus+H$, where $R_\oplus=6\,371$~km being the Earth radius and $H$ being the satellite altitude above the ground.
The slant range between the OGS and the satellite is obtained then from simple trigonometric considerations, [see Fig.~\ref{fig:Geometry1} (a) and Appendix~\ref{app:distance}]
 \begin{align}
\label{eq:LcosZ}
 L(Z)=\sqrt{H^2+2HR_{\oplus}+R_{\oplus}^2\cos^2 Z}-R_{\oplus}\cos Z,
\end{align}
where $Z$ is the zenith angle lying between the vertical direction of the observer and the direction pointing on the satellite.
In Appendix~\ref{app:distance} we show that for the satellite orbits inclined to the observer meridian plane on angle $\Delta\iota$ the zenith angle varies in the range $Z\in[Z_{\mathrm{min}}^{\Delta\iota},\pi/2]$ with
\begin{align}
\label{eq:Zmintext}
Z_{\mathrm{min}}^{\Delta\iota}=\arccos\left[\sqrt{1-\cos^2\Psi\sin^2\Delta\iota}\right],
\end{align}
where $\Psi$ is the geographical latitude of the observer.
Consequently, the slant range (\ref{eq:LcosZ}) varies from the minimal value $L(Z_{\mathrm{min}}^{\Delta\iota})$ at zenith to the maximal value  $\sqrt{(R_\oplus+H)^2-R_\oplus^2}$ at the observer's horizon.

In order to relate the orbit inclination angle $\Delta\iota$ with the characteristics of the relative motion of the satellite and the observer, we restrict our attention to the communication scenario shown in Fig.~\ref{fig:Satellites} for the $\text{OGS}_2$.
If the satellite is initially  at zenith of the observer (indicated as $S_1$ in Fig.~\ref{fig:Satellites}) and moves towards the observer's horizon along the zenith trajectory (i), after one satellite orbiting period $T_{\mathrm{sat}}$  it  reappears at $S_3$  and moves along the inclined trajectory (ii).
The trajectory (ii) is then inclined to the observer's meridian plane due to the Earth rotation.
After $n$th satellite revolution the inclination angle $\Delta\iota$  between the satellite orbit plane and the observer meridian is given by
\begin{align}
\label{eq:iota1}
 \Delta\iota=n\frac{T_{\mathrm{sat}}v_\oplus}{ R_\oplus},
\end{align}
where  $v_\oplus=1669.8$ km/h is the speed of Earth's rotation at the Equator.
We refer the orbit with zero inclination, $\Delta\iota=0$, as to the zenith orbit.
In this case, the orbit plane coincides with the observer meridian.
In our convention, $\Delta\iota$ is positive (negative) if the satellite flies westwards (eastwards) of the observer meridian plane.

In order to get some impression of the communication geometry with the satellite with the inclined orbit, we consider two satellite positions $S$ and $S^\prime$ at successive times and the observer at $O$ as shown in Fig.~\ref{fig:Geometry1} (b).
These positions are chosen in such way that the slant ranges $OS$ and $OS^\prime$ are equal as well as the corresponding zenith angles, i.e., $\sphericalangle SOB=\sphericalangle S^\prime OB=Z$.
Let us choose the plane $ABOC$ as the observer meridian.
Then, the instantaneous position of the satellite is given by both its zenith and azimuth angles.
The latter is the angle between the projected segment that connects the observer with the satellite and the reference vector in the meridian plane pointed toward the north pole.
Clearly, for the inclined orbit at geographical latitudes different from the polar or equatorial ones, the positions $S$ and $S^\prime$ have the same zenith angle but different azimuth angles.

One can imagine that the points $S$ and $S^\prime$ lie on the cross sections of the satellite trajectory with the right circular cone  whose height is aligned along the observer's zenith and the apex coincides with $O$ [cf. the segment of the cone $OASS^\prime$ in Fig.~\ref{fig:Geometry1} (b)].
For inclined orbits that pass above the observer's horizon, each cone of angle $2Z$ crosses the trajectory in two points if $Z\in(Z_{\mathrm{min}}^{\Delta\iota},\pi/2)$ and in one point if $Z=Z_{\mathrm{min}}^{\Delta\iota}$.
Due to this symmetry we are able to characterize the length of geometrical link between the observer and the satellite by Eq.~(\ref{eq:LcosZ}) with $Z\in[Z_{\mathrm{min}}^{\Delta\iota},\pi/2]$ and ignore the detailed information on the corresponding azimuth values.
The latter assumption is well justified if the detailed position of satellite is determined by a tracking system that aligns automatically the sender and the receiver telescopes.
In this case, the observer coordinate system is associated with the plane $OSC$ [cf. Fig.~\ref{fig:Geometry1} (a)] that rotates around the axis $OC$.
Therefore, for optical communication with the automatical azimuth angle tracking the relevant link information is incorporated in the zenith angle dependence of the relevant quantities.
It is worth to note that the cross section of the cone of given $Z$ with other satellite trajectory determines the slant ranges of the same length.
For example, link $AO$ on Fig.~\ref{fig:Geometry1} (b) for the zenith orbit $AB$ is equivalent to the links $SO$ and $S^\prime O$ which results in equal propagation properties of light along these links.


\subsection{Regular refraction}
\label{sec:Refraction}

In this subsection we calculate the elongation of the slant range due to atmospheric refraction.
Atmospheric refraction phenomena are based on the fact that Earth's atmosphere has an optical refractive index that is different from its value in vacuum.
Furthermore, the value of the refractive index varies with the altitude, geographical location, and meteorological conditions and hence refraction depends on space and time variables.
We refer to  refraction phenomena that systematically occur in the atmosphere as to regular refraction~\cite{Mahan1962, Wittmann1997, Young2006, Jiang2014}.
Regular refraction changes with altitude in a theoretically predictable fashion.
Time variation of regular atmospheric refraction has rather seasonal behavior even for large zenith angles ~\cite{Sampson2003} and can be ignored.

Due to a spatial variability of the refractive index with altitude, the light coming from a distant source and reaching the ground  observer propagates along a curved path rather than a straight line.
As a consequence, the signal from distant objects arrives under the apparent zenith angle $Z_a$ rather than under the true zenith angle $Z$.
Both zenith angles are related as
\begin{align}
\label{eq:apparent}
 Z_a=\arcsin\left(\frac{1}{n_0}\sin Z\right)
\end{align}
where $n_0=1.00027$ is the air refractive index near the ground.
We consider the effect of regular refraction on elongation  of the slant range given $L(Z)$.

Earth's atmosphere can be viewed as a spherically stratified medium with specific distribution of refractive index values within each strata.
In order to obtain this distribution we use the so-called standard atmosphere model~\cite{StandardAtmosphere}.
The standard atmosphere is an idealized steady-state representation of Earth's atmosphere that gives values of atmospheric pressure, temperature, and other parameters for altitudes up to 1000 km.
The altitude-dependent values of  the air refractive index  can be found using the distributions of temperature and pressure according to the Edl\'en  equation~\cite{Edlen1966, Birch1994}.
We distinguish 10 atmospheric layers above the ground and within each layer we approximate the latitude dependence of refractive index in linear manner (for details see Appendix~\ref{app:standardAtmosphere})
If we denote the latitude of $i$-th layer upper bound as $H_i$, the linear path within the layer  is determined as
\begin{align}
\label{eq:Lmeridian1}
 L_i&=\Bigl\{(R_\oplus+H_{i-1})^2+(R_\oplus+H_i)^2\\
 &-2(R_\oplus+H_{i-1})(R_\oplus+H_i)\cos[\Phi(Z,r_i)]\Bigr\}^{1/2}.\nonumber
\end{align}
The rest of the total optical path lies in the vacuum and is given by
\begin{align}
\label{eq:Lmeridian2}
 L_{11}&=\Bigl\{(R_\oplus+H_{10})^2+(R_\oplus+H)^2\\
 &-2(R_\oplus+H_{10})(R_\oplus+H)\cos[\Theta(Z, r_{10})]\Bigr\}^{1/2}.\nonumber
\end{align}
The angles $\Phi$ and $\Theta$ determine the relative position of incoming and refracted light rays within each layer and are  complex functions of observer's zenith angle $Z$ and the so-called  refraction integral $r_i$~\cite{Young2006}.
The explicit expressions for the lengths (\ref{eq:Lmeridian1}) and (\ref{eq:Lmeridian2}) are given in Appendix~\ref{app:refr} [cf. Eqs.~(\ref{eq:PhiApp}) and (\ref{eq:ThetaApp})].

It is worth to note that it was assumed that the paths segments $L_i$ in the form (\ref{eq:Lmeridian1}) are obtained with the assumption that they are linear within each atmospheric layer.
This is approximately true for the standard atmosphere model, which gives the minimal ray curvature of $4.4 R_\oplus$ near the ground at $Z_a{=}90^\circ$ (cf. Ref.~\cite{Gaifillia2016} for empirical formulas).
This allows us to neglect the ray curvature effects in the following.
This is of course an idealization and for specific  daytime and meteorological conditions  the accounting for ray curvature effects turns out to be important~\cite{Hirt2010}.

\begin{figure}[ht]
 \includegraphics[width=0.45\textwidth]{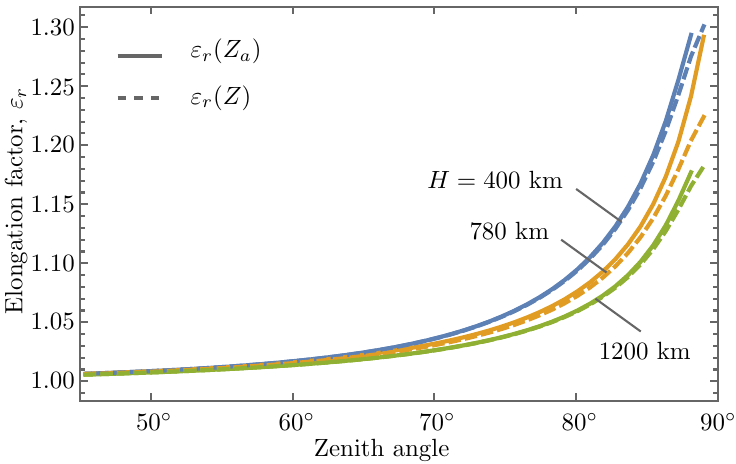}
 \caption{\label{fig:Elongation}
Optical path length elongation factor due to atmospheric refraction as a function of observer's apparent, $Z_a$, and true, $Z$,  zenith angles for several LEO satellite orbits with altitudes $H$. }
\end{figure}

Figure~\ref{fig:Elongation} shows the path elongation factor due to atmospheric refraction,
\begin{align}
\label{elongRefr}
 \varepsilon_r(Z_a)=\frac{1}{L(Z_a)}\sum_{i=1}^{11} L_i(Z_a),
\end{align}
where $L(Z_a)$ is given by Eqs.~(\ref{eq:LcosZ}) and (\ref{eq:apparent}) and $L_i(Z_a)$ are defined by Eqs.~(\ref{eq:Lmeridian1}) and (\ref{eq:Lmeridian2}).
Alternatively, Eq.~(\ref{elongRefr}) can be written as a function of the true zenith angle, i.e. $\varepsilon_r(Z)$.
The behavior of $\varepsilon_r$  as a function of the apparent  zenith angle $Z_a$ starts to deviate from $\varepsilon_r(Z)$ while approaching the horizon due to larger optical density along the propagation path near the horizon.
The elongation factor diminishes with the growth of orbit altitude.
This is due to the increase of the propagation path in the vacuum with the increase of satellite altitude.
For further convenience we give a polynomial fit to the elongation factor as a function of the apparent zenith angle (in degrees) for the orbit with $H=780$ km:
\begin{align}
 \varepsilon_r(Z_a)&=1+1.818908{\times} 10^{-4} Z_a^2\\
 &-4.066061{\times} 10^{-5}|Z_a|^3+3.813573{\times} 10^{-6} Z_a^4\nonumber\\
 &-1.920844{\times}10^{-7}|Z_a|^5+5.710429{\times}10^{-9}Z_a^6\nonumber\\
 &-1.032821{\times}10^{-10}|Z_a|^7+1.117105{\times}10^{-12}Z_a^8\nonumber\\
 &-6.644358{\times}10^{-15}|Z_a|^9+1.672433{\times}10^{-17} Z_a^{10}.\nonumber
\end{align}
Finally, the slant range that accounts the elongation due to atmospheric refraction reads as
\begin{align}
 \label{eq:SlantRefr}
 L_r(Z_a)=\varepsilon_r(Z_a)L(Z_a),
\end{align}
where $L$ is given by Eq.~(\ref{eq:LcosZ}).


\subsection{Regular extinction}
\label{sec:DeterministicExt}

Another source of losses that can be considered as regular or deterministic losses are associated with molecular and aerosol scattering.
For the horizontal atmospheric links,  the extinction factor is $\chi_{\mathrm{ext}}=\exp[-\beta_{\mathrm{ext}}(h)L]$, where $\beta_{\mathrm{ext}}(h)$  is the extinction coefficient due to molecular absorption and scattering at given height $h$ above sea level and $L$ is the distance between the communication parties.
Clearly, for elevated links this formula should be modified in order to include the variation of the extinction coefficient with the height.
If the slant range to the satellite is given by Eq.~(\ref{eq:SlantRefr}) we can write
\begin{align}
\label{eq:chiExt}
 \chi_{\mathrm{ext}}=\exp\Bigl[-\int\limits_0^{L_r(Z_a)}\D L^\prime \beta_{\mathrm{ext}}(L^\prime)\Bigr],
\end{align}
where  the observer is assumed to be located at sea level and $L^\prime=h\sec Z_a$.
Since the value of the extinction coefficient depends on the number density $N(h)$ of air constituents at given height $h$ its altitude dependence can be written as
\begin{align}
\label{eq:betadepend}
  \beta_{\mathrm{ext}}(L^\prime)=\beta_{\mathrm{ext}}^0\frac{N(L^\prime)}{N_0}=\beta_{\mathrm{ext}}^0 \exp\left[-L^\prime/(\mathcal{H}_0\sec Z_a)\right],
\end{align}
where $\beta_{\mathrm{ext}}^0$ is the extinction coefficient at sea level and $N_0$ is the corresponding number density of air constituents.
Here, we have used the altitude dependence of $N$ derived within the standard atmosphere model [cf. Appendix~\ref{app:standardAtmosphere}, Eq.~(\ref{eq:StAtmNdensity})] with the scale parameter $\mathcal{H}_0=6600$ m.
Substituting Eq.~(\ref{eq:betadepend}) in Eq.~(\ref{eq:chiExt}) we finally derive
\begin{align}
\label{eq:overlineL}
 \chi_{\mathrm{ext}}=\exp\Bigl[-\beta_{\mathrm{ext}}^0 \frac{\mathcal{H}_0\sec Z_a}{1000}\left(1-e^{-L_r(Z_a)/(\mathcal{H}_0\sec Z_0)}\right)\Bigr].
\end{align}
Here we have adopted the conventional units of the extinction coefficient to be given in km$^{-1}$.
Molecular or Rayleigh scattering contributes $2.544\times 10^{-3}$ km${^{-1}}$ to this coefficient for optical wavelength $\lambda=800$ nm ~\cite{Elterman1968}.
The aerosol distribution has a more complex dependence on altitude than the model (\ref{eq:StAtmNdensity}).
Nevertheless, for the most part of the optical propagation path one can consider that aerosol scattering contributes roughly the same amount to extinction as the Rayleigh scattering.
For the total extinction coefficient in (\ref{eq:chiExt}), we adopt therefore the value $\beta_{\mathrm{ext}}^0=5\times 10^{-3}$ km${^{-1}}$.
We also note that a similar expression to (\ref{eq:overlineL}) has been derived in Ref.~\cite{Duntley1948}.


\section{Atmospheric turbulence}
\label{sec:Turbulence}

In the previous section we considered the influence of regular refraction on the elongation of the optical path length for light propagating in the atmosphere.
The regular refraction is caused by the variation of the air refraction index with the altitude and is predictable provided the altitude variation of temperature and pressure is known.
Other types of refractive and diffraction phenomena arise due to the irregular variation in space and time of the refractive index.
Such random variations are connected  with temperature  fluctuations and wind shear and have statistical properties of turbulent scalar fields.
The strength of turbulent refractive index fluctuations also varies with the altitude.
Hence, the irregular disturbances of the optical signal in satellite-mediated communication depend on the zenith angle.
In this section we discuss the dependence of aperture-averaged scintillations, beam broadening, and beam wandering  on the  zenith angle.

\subsection{Statistical description of optical turbulence}

Turbulent air motion consists of a set of vortices or eddies of various diameters, ranging from extremely large with characteristic so-called outer scale $L_o$ to extremely small with a scale $l_o$.
Under the influence of inertial forces, larger eddies break up into smaller ones.
This cascade process continues until the minimal scale $l_o$, referred to as the inner scale, is reached and dissipation of turbulent flow energy takes place.
This evolution of turbulent air vortices leads to a random variability of the refractive index,
\begin{align}
\label{eq:nedlen}
 n_i(\boldsymbol{\rho},t)=n_i+\delta n(\boldsymbol{\rho},t),
\end{align}
where $\boldsymbol{\rho}=(x\quad y\quad z)^T$ and  $n_i$ is the regular part of the refractive index within $i$th atmospheric layer [cf.~Eq.~(\ref{eq:refIndHeight}) of Appendix~\ref{app:standardAtmosphere}].
In the following we will omit the index $i$ for convenience and adopt the notation $n(\boldsymbol{\rho},t)$ for the refractive index (\ref{eq:nedlen}).
In Eq.~(\ref{eq:nedlen}) the $z$ axis is chosen along the optical ray while $\boldsymbol{r}{=}(x\quad y)^T$ lies in the transverse plane to the ray direction.
At optical wavelengths, the fluctuating part $\delta n$ is of the order of $10^{-6}$ or less.
When the Taylor's "frozen turbulence" hypothesis~\cite{Taylor1938} holds, implying that the random field $n(\boldsymbol{\rho},t)$ is transported with constant velocity $\boldsymbol{v}$, therefore remaining stationary
in the moving  coordinate system, $n(\boldsymbol{\rho},t)=n({\boldsymbol{\rho}}-{\boldsymbol{v}}t)$, the time dependence of refractive index is incorporated in the spatial variable.

In the statistical theory of optical turbulence, the refractive index fluctuations $\delta n$ are described by the power spectrum
\begin{align}
 \Phi_n(\boldsymbol{k})=\frac{1}{(2\pi)^3}\int\limits_{\mathbb{R}^3}\D^3\boldsymbol{\rho}\, B_n(\boldsymbol{\rho})e^{-i\boldsymbol{k}\cdot\boldsymbol{\rho}},
\end{align}
where the $B_n$ is the correlation function
\begin{align}
 B_n(\boldsymbol{\rho}_1-\boldsymbol{\rho}_2)&=\langle\left[n(\boldsymbol{\rho}_1)-\langle n\rangle\right]\left[n(\boldsymbol{\rho}_2)-\langle n\rangle\right]\rangle\\
 &=\langle\delta n(\boldsymbol{\rho}_1)\delta n(\boldsymbol{\rho}_2)\rangle.\nonumber
\end{align}
For a locally isotropic random field the correlation function simplifies further and the turbulent spectrum can be written as
\begin{align}
\label{eq:PhiTurb}
 &\Phi_n(\boldsymbol{\kappa};z)=\frac{1}{\pi}\int\limits_0^z\D z^\prime F_n(\boldsymbol{\kappa};z^\prime),\\
 & F_n(\boldsymbol{\kappa};z)=\int\limits_{\mathbb{R}^2}\D^2\boldsymbol{r} B_I(\boldsymbol{r},z) e^{-i\boldsymbol{\kappa}\cdot\boldsymbol{r}},
\end{align}
where  $\boldsymbol{\kappa}{=}(k_x\quad k_y)^T$ and $\boldsymbol{r}{=}\boldsymbol{r}_1{-}\boldsymbol{r}_2$.
For the inertial range of spatial wave-numbers values, $|\boldsymbol{\kappa}|\in[ 2\pi/L_o, 2\pi/l_o]$, the   spectrum  for Kolmogorov turbulence is
\begin{align}
\label{eq:Kolmogorov}
 \Phi_n(\boldsymbol{\kappa},z)=0.033 C_n^2(z)|\boldsymbol{\kappa}|^{-\frac{11}{3}}.\qquad
\end{align}
Here, $C_n^2$ is the refractive index structure parameter characterizing the strength of refractive index fluctuations.

In the following considerations we will require the specific form of correlation function, called structure function, for optical phase fluctuations:
\begin{align}
\label{eq:phaseSF}
 \mathcal{D}_S(\boldsymbol{r}_1,\boldsymbol{r}_2,\boldsymbol{r}_1^\prime,\boldsymbol{r}_2^\prime;z)=\langle\left[S(\boldsymbol{r}_1,\boldsymbol{r}_1^\prime;z)-S(\boldsymbol{r}_2,\boldsymbol{r}_2^\prime;z)\right]^2\rangle,
\end{align}
where  $\boldsymbol{r}_i$, $\boldsymbol{r}_i^\prime$, $i=1,2$, are the transverse components of the corresponding spatial vectors $\boldsymbol{\rho}_i=(\boldsymbol{r}_i\quad z)$ and $\boldsymbol{\rho}_i^\prime=(\boldsymbol{r}_i^\prime\quad z)$.
Here, the fluctuating phase is found from the first approximation of geometric optics
\begin{align}
\label{eq:SUL}
 S^{\mathrm{UL}}(\boldsymbol{r},\boldsymbol{r}^\prime;z)=\frac{k}{2}\int\limits_0^{z}\D z^\prime\,\delta n\left(\boldsymbol{r}\frac{z^\prime}{z}+\boldsymbol{r}^\prime\frac{z-z^\prime}{z},z^\prime\right)
\end{align}
for uplink  and
\begin{align}
\label{eq:SDL}
 S^{\mathrm{DL}}(\boldsymbol{r},\boldsymbol{r}^\prime;z){=}\frac{k}{2}\int\limits_0^{z}\D z^\prime\,\delta n\left(\boldsymbol{r}\frac{z-z^\prime}{z}{+}\boldsymbol{r}^\prime\frac{z^\prime}{z},L{-}z^\prime\right)
\end{align}
for downlink  communication scenarios.
In this article we focus our attention on the downlink communication configuration and omit the corresponding superscripts for simplicity in the notations.
The formulas for uplink can be then obtained by replacing $z^\prime\rightarrow L-z^\prime$ in integrals similar to the one in Eq.~(\ref{eq:SDL}).

For a locally isotropic and homogeneous random field $\delta n$, the Markov approximation,
\begin{align}
 \langle\delta n(\boldsymbol{r};z)\delta n(\boldsymbol{r}^\prime;z^\prime)\rangle&=2\pi\delta(z-z^\prime)\int\limits_{\mathbb{R}^2}\D^2\boldsymbol{\kappa}\nonumber\\
 &\times \Phi_n(\boldsymbol{\kappa},z)\exp\left[i\boldsymbol{\kappa}\cdot(\boldsymbol{r}-\boldsymbol{r}^\prime)\right],
\end{align}
is well justified \cite{Tatarskii2016, Baskov}.
The phase structure function (\ref{eq:phaseSF}) is evaluated then as
\begin{align}
\label{eq:StrFunctS}
 &\mathcal{D}_S(\boldsymbol{r}_1,\boldsymbol{r}_2,\boldsymbol{r}_1^\prime,\boldsymbol{r}_2^\prime;z)=\mathcal{D}_S(\boldsymbol{r}_1-\boldsymbol{r}_2,\boldsymbol{r}_1^\prime-\boldsymbol{r}_2^\prime;z)\nonumber\\
 &=\pi k^2\int\limits_0^{z}\D z^\prime\int\limits_{\mathbb{R}^2}\D^2\boldsymbol{\kappa}\Phi_n(\boldsymbol{\kappa},z{-}z^\prime)\\
 &\times\Biggl\{1-\exp\Bigl(i\boldsymbol{\kappa}{\cdot}\Bigl[(\boldsymbol{r}_1{-}\boldsymbol{r}_2)\frac{z-z^\prime}{z}+(\boldsymbol{r}_1^\prime{-}\boldsymbol{r}_2^\prime)\frac{z^\prime}{z}\Bigl]\Bigr)\Biggr\}.\nonumber
\end{align}
For the Kolmogorov spectrum (\ref{eq:Kolmogorov}), in the case of downlink the structure function (\ref{eq:StrFunctS}) reduces to \cite{Gurvich}
\begin{align}
\label{eq:DsKolmogorov}
 \mathcal{D}_S&(\boldsymbol{r},\boldsymbol{r}^\prime,L_r)\\
 &=2\rho_0^{-\frac{5}{3}}\int\limits_{0}^1\D\xi \frac{C_n^2([1{-}\xi] L_r)}{C_{n,0}^2}\left|\boldsymbol{r}(1-\xi){+}\boldsymbol{r}^\prime\xi\right|^{\frac{5}{3}},\nonumber
\end{align}
where
\begin{align}
\label{eq:rho0}
 \rho_0\approx(1.5 C_{n,0}^2k^2 L_{\mathrm{turb}})^{-3/5}
\end{align}
is the radius of spatial coherence of a plane wave in the atmosphere, $L_{\mathrm{turb}}$ is the propagation length within  the optically active turbulent atmospheric layer, $L_r$ is the total propagation length, $\xi$ is the dimensionless integration variable, and
\begin{align}
\label{eq:Cn0}
 C_{n,0}^2=C_n^2(h_0\sec Z_a)
\end{align}
 is the refractive index structure function taken at some reference height $h_0$ above the ground (see also Appendix~\ref{app:Cn2models}).
The structure function (\ref{eq:DsKolmogorov}) for uplink is obtained by the change of the variable $\xi\rightarrow 1-\xi$.

\subsection{Aperture-averaged scintillations}
\label{sec:SciRigorous}

The dependence of optical intensity fluctuations, i.e., scintillations, on the zenith angle has attracted attention in the connection with astronomical photometry.
Early measurements of the scintillation index $\sigma_\eta^2$ \cite{Butler1952, Ellison1952, Briggs1963}  have shown that the aperture-averaged scintillations (or power scintillations) grow with the growth of the zenith angle as
\begin{align}
\label{eq:scintEta}
 \sigma_\eta^2=\frac{\langle\Delta \eta^2\rangle}{\langle \eta\rangle^2}\propto\left(\sec Z_a\right)^\gamma,
\end{align}
where
\begin{align}
\label{eq:eta}
 \eta=\int_{\mathcal{A}}\D^2 \boldsymbol{r}I(\boldsymbol{r}, L_r)
\end{align}
is the transmittance of the light intensity $I(\boldsymbol{r}, L_r)$ through the receiver aperture with opening area $\mathcal{A}$, and $\boldsymbol{r}$ is the spatial variable transversal to propagation direction.
The exponent $\gamma$ in Eq.~(\ref{eq:scintEta}) is related to the statistics of turbulent fluctuations of the refractive index and depends on the characteristics of the receiver telescope.
Theoretical considerations~\cite{Tatarskii2016} based on the Rytov approximation yield $\gamma=11/6$ for small receiving apertures and $\gamma=3$ for large receiving apertures, respectively.
These results agree reasonably with the experiments~\cite{Stecklum1985}.

The later investigations~\cite{Burke1970, Stecklum1985, Kucherov, Fuentes1987} have shown that dependence (\ref{eq:scintEta}) is valid for small and moderate zenith angles or for highly elevated optical ground stations.
For large zenith angles measured stellar scintillations exhibit a saturation or  decrease in its value.
A similar behavior has been reported for optical signals from satellites~\cite{Moll2014, Moll2015}.
On the other hand, the Rytov approximation (\ref{eq:scintEta}) yields the divergent scintillation index for $Z_a{\rightarrow}90^\circ$ and does not account for the saturation effect.
The phenomenon of saturation of scintillations has been theoretically studied in Refs.~\cite{Young1969, Young1970, Young1970a, Beran1988, Andrews} and is attributed to multiple scattering phenomena on turbulent inhomogeneities.

In this section we derive the scintillation index of the aperture-averaged optical signal from the satellite and discuss its dependence on the zenith angle.
As an accompanying result we obtain the expressions for two first moments of the transmittance (\ref{eq:eta}), the mean width of the beam spot at the receiver and the beam wandering variance.
These parameters will be used in Sec.~\ref{sec:PDT} for the derivation of the probability distribution of quantum channel transmittance.

The scintillation index (\ref{eq:scintEta}) is derived from the first two moments of the transmittance (\ref{eq:eta}), namely
\begin{align}
\label{eq:eta1}
 \langle\eta\rangle=\int\limits_{|\mathbf{r}|\le a}\D^2\boldsymbol{r}\,\Gamma_2(\boldsymbol{r};L_r),
\end{align}
\begin{align}
\label{eq:eta2}
 \langle\eta^2\rangle=\int\limits_{|\mathbf{r}_1|\le a}\D^2\boldsymbol{r}_1\int\limits_{|\mathbf{r}_2|\le a}\D^2\boldsymbol{r}_2\,\Gamma_4(\boldsymbol{r}_1\boldsymbol{r}_2;L_r),
\end{align}
where
\begin{align}
\label{eq:Gamm2}
 \Gamma_2(\boldsymbol{r};z)=\langle u(\boldsymbol{r};z)u^\ast(\boldsymbol{r};z)\rangle=\langle I(\boldsymbol{r};z)\rangle, %
\end{align}
\begin{align}
\label{eq:Gamm4}
\Gamma_4(\boldsymbol{r}_1,\boldsymbol{r}_2;z)&=\langle u(\boldsymbol{r}_1;z)u^\ast(\boldsymbol{r}_1;z)u(\boldsymbol{r}_2;z)u^\ast(\boldsymbol{r}_2;z)\rangle\nonumber\\
&=\langle I(\boldsymbol{r}_1;L)I(\boldsymbol{r}_2;z)\rangle
\end{align}
$a$ is the receiving aperture radius, and $u(\boldsymbol{r};L)$ is the optical field amplitude.
Obviously, the field correlation functions $\Gamma_2$ and $\Gamma_4$ turn to be important ingredients for the evaluation of  $\sigma_\eta^2$.
On the other hand, their moments are related to the beam wandering variance,
\begin{align}
\label{eq:BW}
 \sigma_{\mathrm{BW}}^2=\int\limits_{\mathbb{R}^4}\D^2\boldsymbol{r}_1\D^2\boldsymbol{r}_2\,x_1 x_2 \Gamma_4(\boldsymbol{r}_1,\boldsymbol{r}_2;L),
\end{align}
and to the mean short-term beam spot radius of the transmitted beam,
\begin{align}
\label{eq:WST}
 W_{\mathrm{ST}}&=\sqrt{W_{\mathrm{LT}}^2-4\sigma_{\mathrm{BW}}^2}\\
 &=2\Biggl[\,\int\limits_{\mathbb{R}^2}\D^2\boldsymbol{r} x^2\Gamma_2(\boldsymbol{r};L)-\sigma_{\mathrm{BW}}^2\Biggr]^{1/2}.\nonumber
\end{align}
Here the variable $x$ denotes the $x$ component of the transverse vector $\boldsymbol{r}$.
The short-term beam spot radius  is associated with the intensity distribution observed during small exposure times while the long-term radius $W_{\mathrm{LT}}$ includes broadening effects due to beam wandering and is associated with long detection times.
In the following section we show that the quantities (\ref{eq:eta1}), (\ref{eq:eta2}), (\ref{eq:BW}), (\ref{eq:WST}) are of primary importance for the description of quantum atmospheric channels.

The correlation functions $\Gamma_2$ and $\Gamma_4$ we calculate using the phase approximation of the Huygens-Kirchhoff method~\cite{Banakh1977, Banakh1979}.
This approximative method neglects the fluctuations of the  field amplitude.
On the other hand, the  phase fluctuations characterized by Eqs.~(\ref{eq:SUL}) and (\ref{eq:SDL}) are caused by random diffraction and refraction in the turbulent medium, which are included up to the terms of order $\langle\delta n^2\rangle^{1/2}$.
The former condition is justified for  communication scenarios through links with saturated turbulence ~\cite{Holmes1980}, which is the case for satellite-mediated atmospheric channels.
More rigorous methods \cite{Klyatskin1974} and experiments \cite{Tatarskii2016, Artemev1971} suggest that the  latter condition is satisfied for arbitrary propagation paths and turbulence strengths.
Thus, the phase approximation of the Huygens-Kirchhoff method accounts  for the diffraction and refraction effects arising due to the optical beam propagation in the turbulent atmosphere in most practically important cases.

For the TE00 mode of the laser beam  in the plane of the radiating aperture,
\begin{align}
 u(\boldsymbol{r};0)=\sqrt{\frac{2}{\pi^2W_0^2}}\exp\left[-\left(\frac{1}{W_0^2}-\frac{ik}{2 F}\right)r^2\right],
\end{align}
with $W_0$ being the initial beam spot size  and $F$ being the initial beam wavefront radius, the correlation functions (\ref{eq:Gamm2}) and (\ref{eq:Gamm4}) read as
\begin{align}
\label{eq:Gamma2HK}
 \Gamma_2(\boldsymbol{r};L_r)&{=}\frac{k^2}{4\pi^2L_r^2}\int\limits_{\mathbb{R}^2}\D^2\boldsymbol{r}^\prime \\ 
 &\times e^{-\frac{g^2 {r^\prime}^2}{2W_0^2}-2i\frac{\Omega}{W_0^2}\boldsymbol{r}\cdot\boldsymbol{r}^\prime-\frac{1}{2}\mathcal{D}_S(0,\boldsymbol{r}^\prime;L_r)}\nonumber,
\end{align}
\begin{align}
\label{eq:Gamma4HK}
 &\Gamma_4(\boldsymbol{r},\boldsymbol{R};L_r)=\frac{2 k^4}{\pi^2(2\pi)^3L_r^4 W_0^2}\int\limits_{\mathbb{R}^6}\D^2\boldsymbol{r}_1^\prime \D^2\boldsymbol{r}_2^\prime\D^2\boldsymbol{r}_3^\prime\nonumber\\
 &\times 
 \,e^{-\frac{1}{W_0^2}({r_1^\prime}^2+{r_2^\prime}^2+g^2{r_3^\prime}^2)+2i\frac{\Omega}{W_0^2}[1-\frac{L}{F}]\boldsymbol{r}_1^\prime\cdot\boldsymbol{r}_2^\prime}\\
 &\quad\times e^{-2i\frac{\Omega}{W_0^2}\boldsymbol{r}\cdot\boldsymbol{r}_2^\prime-4i\frac{\Omega}{W_0^2}\boldsymbol{R} \cdot\boldsymbol{r}_3^\prime}\mathcal{J}(\boldsymbol{r},\boldsymbol{r}_1^\prime,\boldsymbol{r}_2^\prime,\boldsymbol{r}_3^\prime)\nonumber
\end{align}
\begin{align}
\label{eq:Jkernel}
 &\mathcal{J}=\exp\Bigl[\frac{1}{2}\sum\limits_{j=1,2}\Bigl\{\mathcal{D}_S(\boldsymbol{r},\boldsymbol{r}_1^\prime+(-1)^j\boldsymbol{r}_2^\prime;L_r)\\
 &-\mathcal{D}_S(\boldsymbol{r},\boldsymbol{r}_1^\prime{+}(-1)^j\boldsymbol{r}_3^\prime;L_r){-}\mathcal{D}_S(0,\boldsymbol{r}_2^\prime{+}(-1)^j\boldsymbol{r}_3^\prime;L_r)\Bigl],\nonumber
\end{align}
where we have used the relative and the center-of-mass coordinates $\boldsymbol{r}=\boldsymbol{r}_1-\boldsymbol{r}_2$ and $\boldsymbol{R}=(\boldsymbol{r}_1+\boldsymbol{r}_2)/2$, respectively.
Here, $\Omega=kW_0^2/(2L_r)$ is the Fresnel number of the transmitter aperture, $g^2=1+\Omega^2[1-L_r/F]^2$ is the generalized diffraction parameter, and the phase structure function $\mathcal{D}_S$ is given by (\ref{eq:DsKolmogorov}).
We also note that Eqs.~(\ref{eq:Gamma2HK}) and (\ref{eq:Gamma4HK}) incorporate multiple-scattering effects.

\subsubsection{Mean transmittance}

The average intensity at the receiver aperture is found from Eqs.~(\ref{eq:Gamm2}) and (\ref{eq:Gamma2HK}).
We use the quadratic approximation~\cite{Andrews2005}
\begin{align}
\label{eq:quadrAppr}
 \exp\left[-\left(r/\rho_0\right)^{\frac{5}{3}}\right]\approx\exp\left[-\left(r/\rho_0\right)^2\right],
\end{align}
for the spatial dependence of the phase structure function (\ref{eq:DsKolmogorov}).
This approximation gives a good accuracy for small values of the radius of spatial coherence $\rho_0$, i.e., for long propagation distances and strong optical turbulence, which is always the case for satellite-mediated links.
Performing the integration in Eq.~ (\ref{eq:Gamma2HK}), we obtain the Gaussian distribution of the intensity distribution at the receiver aperture plane,
\begin{align}
\label{eq:Ieff}
 \langle I(\boldsymbol{r}; L_r)\rangle=\frac{2}{\pi W_{\mathrm{LT}}^2}\exp\left[-2\frac{r^2}{W_{\mathrm{LT}}^2}\right], 
\end{align}
where the long-term beam spot radius at the receiver is [cf. also~Eq.~(\ref{eq:WST})]
\begin{align}
\label{eq:WLT}
 W_\mathrm{LT}( L_r)=W_0\left[\left(1-\frac{ L_r}{F}\right)^2+\Omega^{-2}\left(1+ \frac{W_0^2}{\rho_0^2}\mathcal{X}^2\right)\right]^{1/2}.
\end{align}
Here, $\rho_0$ is defined in Eq.~(\ref{eq:rho0}) and
\begin{align}
\label{eq:XCn2}
\mathcal{X}^2=\frac{1}{C_{n,0}^2}\int\limits_0^{1}\D \xi\,C_n^2(L_r,1-\xi) \xi^{5/3}
\end{align}
is the weighting factor that depends on the slant profile of the refractive index structure function $C_n^2(L_r,1-\xi)$.
This profile is related with the vertical profile of turbulence as $C_n^2(L_r,1{-}\xi)=C_n^2[h(\xi){=}L_r(1-\xi)\cos Z_a]$ (see Appendix~\ref{app:Cn2models}~for details).
The natural diffraction of the laser beam is included in Eq.~(\ref{eq:WLT}) by considering the initial beam waist $W_0$, wavelength $\lambda$ (Fresnel number $\Omega$), and propagation distance $L_r$.
The resulting natural laser beam divergence is  usually  the main source of signal loss.
The obtained formula (\ref{eq:WLT}) coincides with the long-term beam spot radius given in Ref.~\cite{Fante1975}.

The obtained mean intensity allows one to calculate the mean transmittance straightforwardly.
Inserting Eq.~(\ref{eq:Ieff}) in Eq.~(\ref{eq:eta}) and performing the integration with respect to the spatial variable, we obtain
\begin{align}
 \langle\eta\rangle\approx1- e^{-\frac{2 a^2}{W_{\mathrm{LT}}^2}}.
\end{align}
This is the  transmittance of Gaussian beam with the beam-spot radius $W_{\mathrm{LT}}$ through a circular aperture of radius $a$.
The approximation sign is used due to the used quadratic approximation (\ref{eq:quadrAppr}).
This formula can serve for estimative calculations only and for the precise evaluations  from Eqs.~(\ref{eq:eta1}) and (\ref{eq:Gamma2HK}) we derive
\begin{align}
\label{eq:etamean}
\langle\eta\rangle{=}\frac{k^2}{L_r^2}\int\limits_0^a\D r r\int\limits_0^\infty\D r^\prime r^\prime  e^{-\frac{g^2 {r^\prime}^2}{2 W_0^2}-\frac{1}{2}\mathcal{D}_S(0,r^\prime;L_r)}J_0\left(\frac{2\Omega}{W_0^2}r r^\prime\right),
\end{align}
where $J_n(x)$ is the Bessel function of $n$th order.

For the sake of completeness, we give the expression of the short-time beam spot size (\ref{eq:WST}).
Following Ref.~\cite{Yura1973} it is calculated as
\begin{align}
\label{eq:WST1}
  W_\mathrm{ST}( L_r)&=W_0\Biggl[\left(1-\frac{ L_r}{F}\right)^2\\
  &+\Omega^{-2}\Biggl(1+ \frac{W_0^2}{\rho_0^2}\frac{\mathcal{X}^2}{1+0.24\left(\frac{\rho_0}{a\mathcal{X}}\right)^{1/3}} \Biggr)\Biggr]^{1/2}.\nonumber
\end{align}
Here, the term proportional to $\mathcal{X}$ arises due to the diffraction-induced beam broadening caused by the turbulent atmosphere.
In the limit $\mathcal{X}\rightarrow 0$ we obtain from Eq.~(\ref{eq:WST1}) the diffraction-induced beam broadening in vacuum.
We also note that Eq.~(\ref{eq:WST1}) does not account for the beam broadening due to the random scattering on particles, aerosols, and precipitations.
This additional broadening might be relevant under the condition of high moisture or low visibility, as has been shown in Ref.~\cite{Vasylyev2017}.
In this article, however, we deal with ideal weather conditions for optical communication.
The short-term beam spot radius~(\ref{eq:WST1}) is important for the derivation of  the probability distribution of the channel transmittance as will be shown in the next section.

\subsubsection{Second moment of transmittance}

The second moment of transmittance is obtained by substituting Eq.~(\ref{eq:Gamma4HK}) in (\ref{eq:eta2}).
Unfortunately, the integration cannot be performed in analytic form.
However, for satellite-mediated atmospheric links we can simplify the integral kernel (\ref{eq:Jkernel}) for the further numerical integration (for details see Appendix~\ref{app:Kernel}):
\begin{widetext}
\begin{align}\label{expans}
&\mathcal{J}(\mathbf{r},\mathbf{r}_1^\prime,\mathbf{r}_2^\prime,\mathbf{r}_3^\prime)\nonumber\\
&\approx\exp\Bigl[-\rho_0^{-\frac{5}{3}}\int\limits_0^1\!\!\D\xi\frac{C_n^2(L_r,1-\xi)}
{C_{n,0}^2}\sum\limits_{j=1,2}\left|\mathbf{r}(1{-}\xi){+}[\mathbf{r}_1^\prime{+}(-1)^j\mathbf{r}_3^\prime]\xi\right|^{\frac{5}{3}}\Bigr]+\exp\Bigl[-\rho_0^{-\frac{5}{3}}\int\limits_0^1\!\!\D\xi\frac{C_n^2(L_r,1-\xi)}{C_{n,0}^2}\sum_{j=1,2}\left|[\mathbf{r}_2^\prime{+}(-1)^j\mathbf{r}_3^\prime]\xi\right|^{\frac{5}{3}}\Bigr]\nonumber\\
&\quad-\exp\Bigl[-\rho_0^{-\frac{5}{3}}\int\limits_0^1\D\xi \frac{C_n^2(L_r,1-\xi)}{C_{n,0}^2}\sum\limits_{j=1,2}\Bigl\{ \left|[\mathbf{r}_2^\prime{+}(-1)^j\mathbf{r}_3^\prime]\xi\right|^{\frac{5}{3}}+\left|\mathbf{r}(1-\xi){+}[\mathbf{r}_1^\prime{+}(-1)^j\mathbf{r}_3^\prime]\xi\right|^{\frac{5}{3}}\Bigr\}\Bigr],
\end{align}
\end{widetext}
This expression can be simplified further by noting that  most of the optical propagation path lies in vacuum as well as in the atmosphere with negligibly small turbulence, i.e., $L_{\mathrm{turb}}\ll L_r$.
As a consequence, for small values of the integration variable $\xi$ in (\ref{expans}), when the contributions from the $\boldsymbol{r}(1-\xi)$ term are dominant, the value of the refractive index structure constant is equal to zero.
On the other hand, for $\xi\approx 1$  the structure constant is finite but the contribution from  $\boldsymbol{r}(1-\xi)$ is negligibly small.
Therefore, one can neglect the dependence on $\boldsymbol{r}$ in (\ref{expans}) along the whole propagation path
\begin{align}
\label{eq:Japprox}
 \mathcal{J}(\mathbf{r},\mathbf{r}_1^\prime,\mathbf{r}_2^\prime,\mathbf{r}_3^\prime)\approx\mathcal{J}(0,\mathbf{r}_1^\prime,\mathbf{r}_2^\prime,\mathbf{r}_3^\prime).
\end{align}
We stress that this formula is justified for downlink  configurations only.
Inserting (\ref{eq:Gamma4HK}) in (\ref{eq:eta2}) and using the approximation (\ref{eq:Japprox}), we obtain
\begin{align}
\label{eq:Gamma4HKapprox}
 &\langle\eta^2\rangle{=}\frac{2 k^4}{\pi^2(2\pi)^3L_r^4 W_0^2}\!\int\limits_{|\boldsymbol{r}_1|\le a}\!\D^2\boldsymbol{r}_1\! \int\limits_{|\boldsymbol{r}_2|\le a}\!\D^2\boldsymbol{r}_2 \int\limits_{\mathbb{R}^6}\D^2\boldsymbol{r}_1^\prime \D^2\boldsymbol{r}_2^\prime\D^2\boldsymbol{r}_3^\prime\nonumber\\
 &\times
 \,e^{-\frac{1}{W_0^2}({r_1^\prime}^2+{r_2^\prime}^2+g^2{r_3^\prime}^2)+2i\frac{\Omega}{W_0^2}[1-\frac{L}{F}]\boldsymbol{r}_1^\prime\cdot\boldsymbol{r}_2^\prime}\\
 &\quad\times e^{-2i\frac{\Omega}{W_0^2}\boldsymbol{r}_1\cdot(\boldsymbol{r}_2^\prime{+}\boldsymbol{r}_3^\prime)+2i\frac{\Omega}{W_0^2}\boldsymbol{r}_2 \cdot(\boldsymbol{r}_2^\prime{-}\boldsymbol{r}_3^\prime)}\mathcal{J}(0,\boldsymbol{r}_1^\prime,\boldsymbol{r}_2^\prime,\boldsymbol{r}_3^\prime).\nonumber
\end{align}
The further evaluation of integrals in (\ref{eq:Gamma4HKapprox}) should be performed numerically.

\begin{figure}[ht]
 \includegraphics[width=0.46\textwidth]{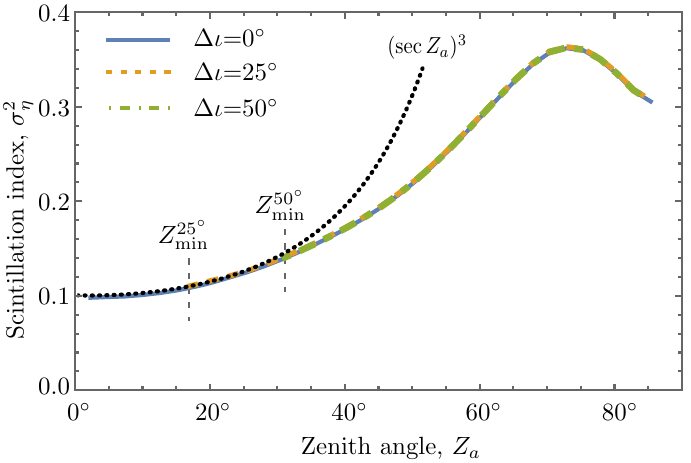}
 \caption{\label{fig:SciRigorous}
Aperture-averaged scintillation index as a function of the zenith angle for several inclination angles of the satellite orbit.
The dotted intervals indicate the minimal zenith angles  for inclined orbits [cf.~Eq.~(\ref{eq:Zmintext})].
The dotted curve corresponds to the asymptotic value (\ref{eq:scintEta}). }
\end{figure}

\begin{table}[h]
\caption{\label{tab:AtmParameters}
Atmospheric, optical beam, and geographical parameters used in the simulations.
Zenith-dependent values are given at $Z_a=0^\circ$.
}
\begin{ruledtabular}
\begin{tabular}{ l c c }
\textrm{Parameter}&\textrm{Notation}&\textrm{Value}\\
\colrule
Radius of spatial coherence    & $\rho_0$  (cm)                &  $13$             \\
Reference structure parameter  & $C_{n,0}^2$ (m$^{-2/3}$)       &  $10^{-17}$        \\
Wavelength                     & $\lambda$ (nm)                &  $840$             \\
Initial beam-spot radius       & $W_0$ (cm)                    &  $2$             \\
Receiver's aperture radius     & $a$ (m)                       &  $0.5$             \\
Wave-front radius              & $F$ (m)                      &  $10^5$              \\
Observer's coordinates          &                           &  $48^\circ$N, $11.5^\circ$~E
\end{tabular}
\end{ruledtabular}
\end{table}

Figure~\ref{fig:SciRigorous} shows the aperture-averaged scintillation index $\sigma_\eta^2=\langle\Delta\eta^2\rangle/\langle\eta\rangle^2$ as a function of the zenith angle calculated by using the phase approximation of the Huygens-Kirchhoff method.
For the calculation of $\langle\eta^2\rangle$ we have used the approximate expression (\ref{eq:Gamma4HKapprox}), whereas  $\langle\eta\rangle$ is calculated from Eq.~(\ref{eq:etamean}).
Table~\ref{tab:AtmParameters} lists the atmospheric and optical beam parameters used in the calculation of the scintillation index.
The curves shown for three inclination angles yield the same functional dependence on zenith angle for medium and large $Z_a$ due to the properties discussed in Sec.~\ref{sec:distance} and differ on the minimal value of the zenith angle $Z_{\mathrm{min}}^{\Delta\iota}$ [ see Eq.~(\ref{eq:Zmintext})].
For small and moderate zenith angles the scintillation index $\sigma_\eta^2$ shows the   asymptotic behavior given by Eq.~(\ref{eq:scintEta}).
The scintillation index calculated within the phase approximation of Huygens-Kirchhoff method shows the  saturation and decrease  of intensity fluctuations for large zenith angles.

\begin{figure}
 \includegraphics[width=0.45\textwidth]{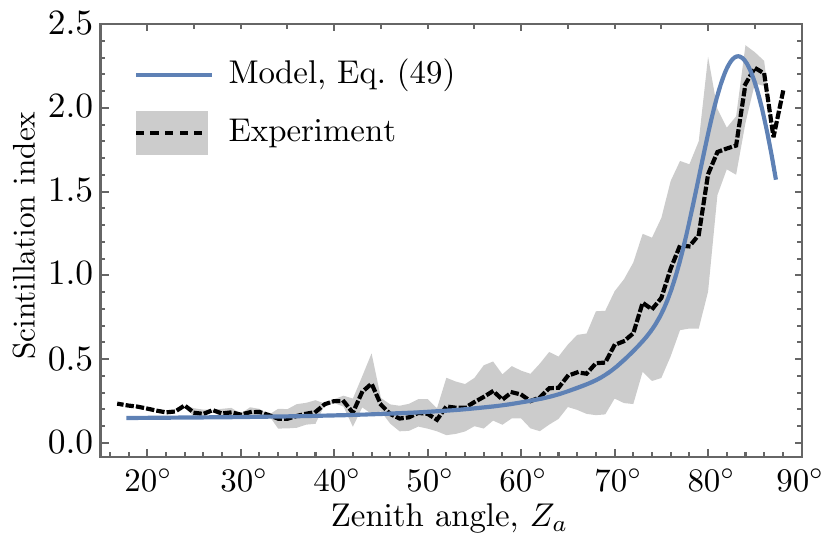}
 \caption{\label{fig:Scintillation}
Intensity scintillation index of LEO-ground  downlink at $847$ nm.
Experimental data are given with confidence intervals (shaded area) and mean value (dashed line) calculated from the individual measurements.
The theoretical curve (solid) is calculated using Eq.~(\ref{eq:sigmaEta2text}).}
\end{figure}

This result can be qualitatively compared with the experimental data that have been taken within five measurement campaigns between 2006 and 2016 and shown in Fig.~\ref{fig:Scintillation} (shaded area).
The direct quantitative comparison of theoretical curves in Fig.~\ref{fig:SciRigorous} with any individual experimental curve  contributing to Fig.~\ref{fig:Scintillation} is hardly possible due to the lack of all needed parameters of turbulence and the hardly measurable profile of $C_n^2(z)$ inherent to the given meteorological conditions.
Measurements of intensity scintillation index  were conducted with an optical ground station nearby Munich, Germany, with elevation of about 602~m above sea level.
Measurement wavelength was 847 and 1550 nm, depending on the used satellite (see Table \ref{tab:Experiment} for an overview of measurement campaigns).

We also note that the difference in zenith angles with maximal  scintillation index in Figs.~\ref{fig:SciRigorous} and \ref{fig:Scintillation} arises due to the different elevations of the observers.

\begin{table}[h]
\caption{\label{tab:Experiment}%
Overview of conducted measurement campaigns.
}
\begin{ruledtabular}
\begin{tabular}{ l c c c }
\textrm{Year}&\textrm{Satellites/}&
\textrm{Wavelength (nm)}&\textrm{No. Measurements}\\
&\textrm{(Laser terminal)}&\\
\colrule
2006  & OICETS (LUCE)  & 847  & 5   \\
2009  & OICETS (LUCE)  & 847  & 4 \\
2015  & ISS (OPALS)   & 1550  & 1 \\
2016  & ISS (OPALS)   & 1550  & 1 \\
2016  & Socrates (SOTA)   & 1550  & 1 \\
\end{tabular}
\end{ruledtabular}
\end{table}

The measurement device that delivered the data for the scintillation index analysis is an infrared camera located in the exit pupil of the telescope.
Thus, images of the intensity field incident onto the telescope aperture are recorded and analyzed.
The experimental curves in Fig.~\ref{fig:Scintillation} are obtained based on data from a single camera pixel.
 The effective radius of the pixel, taking magnification of the optical system into account, is $a=3.2$~mm.
The small value of the detector aperture as well as short exposure times (of order 0.1-1~ms) reduce the telescope aperture smoothing effect, yielding the intensity scintillation index
\begin{align*}
\sigma_I^2=\frac{\langle\Delta I^2\rangle}{\langle I\rangle^2}\approx \sigma_\eta^2\bigr|_{a\rightarrow 0}.
\end{align*}
For the sake of better comparison, the 1550-nm data are re-calculated to 847-nm wavelength using weak scattering theory.
This is possible since the 1550 nm measurements lie well within the weak scattering regime, i.e. at low zenith angles.
Twelve measurements are analyzed to form the mean run of scintillation index as shown in Fig.~\ref{fig:Scintillation}.
The grey area indicates the confidence bound defined by the standard deviation.
At low and high zenith angles, the standard deviation is not illustrated since only a single measurement track is recorded in these regimes and, thus, determination of standard deviation is not possible.
Further description of the individual measurement campaigns and the data analysis method are found in \cite{Moll2015, Moll2016}.

In Appendix~\ref{app:Young} we consider an estimation for the aperture-averaged scintillation index based on approximate phenomenological expressions for field correlation functions.
This can give some insight into saturation effects observed at large zenith angles.
The consideration yields
\begin{align}
\label{eq:sigmaEta2text}
 \sigma_\eta^2&=  1.12\,C_{n,0}^2\,[\Delta\kappa]^{\frac{7}{3}}(H_0\sec Z_a)^3\\
 &\qquad\times{}_2F_3\left(\frac{7}{6},\frac{3}{2};2,\frac{13}{6},3;-a^2\Delta\kappa^2\right),\nonumber\\
 &\Delta\kappa{=}0.69\,\mu\,C_{n,0}^{-6/5}k^{-1/5}(H_0\sec Z_a)^{-8/5}.\nonumber
\end{align}
Here, $a$ is the receiving aperture radius, $k=2\pi/\lambda$ is the optical wave number,  and ${}_2F_3(a,b;c,d,e;x)$ is the hypergeometric function.
This simple analytic formula contains three phenomenological parameters: the refractive index structure parameter at the ground $C_{n,0}^2$,  the characteristic height of the atmospheric turbulent layer $H_0$, and the dimensionless proportionality parameter $\mu$.
The theoretical curve based on Eq.~(\ref{eq:sigmaEta2text}) shows a reasonable agreement with the experimental data in Fig.~\ref{fig:Scintillation}.
The  model parameters are:   $C_{n,0}^2=2.5{\times}10^{-17}$~m$^{-2/3}$,  $H_0=0.5$~km, $a=3.2$~mm, $\lambda=847$~nm, and $\mu=0.92$.

\subsubsection{Beam wandering variance}

For the sake of completeness, we give the expression for the beam wandering variance~(\ref{eq:BW}).
Beam wandering phenomenon depends strongly  on the outer scale of turbulence, $L_o$.
The finite $L_o$ defines the upper bound on the size of turbulent inhomogeneities that are able to deflect the beam as a whole.
Since the Kolmogorov   spectrum (\ref{eq:Kolmogorov}) has a discontinuity at the turbulent wave numbers, $|\boldsymbol{\kappa}|\approx\kappa_o=2\pi/L_o$, it is more suitable to use the smoothed spectrum
\begin{align}
\label{eq:PhiSpectrum}
 \Phi_n(\boldsymbol{\kappa},z)=0.033 C_n^2(z)|\boldsymbol{\kappa}|^{-\frac{11}{2}}\Bigl(1+e^{-\frac{|\boldsymbol{\kappa}|^2}{\kappa_\circ^2}}\Bigr).
\end{align}
For slant paths the outer turbulence scale varies with the height $h$.
Very well known is the empirical Coulman-Vernin profile that reads as~\cite{Coulman1988}
\begin{align}
\label{eq:Lout}
 L_\circ(h)=\frac{4}{1+\Bigl(\frac{h-8500}{2500}\Bigr)^2},
\end{align}
where the outer scale is given in meters.

Using the spectrum (\ref{eq:PhiSpectrum}) and  following the derivation steps of Ref.~\cite{Mironov1977}
we obtain
\begin{align}
\label{eq:BWnew}
 &\sigma_{\mathrm{BW}}^2=1.29 L_r^3\int\limits_0^1\D\xi\xi^2 C_n^2(L_r,1{-}\xi)\Bigl\{W_{\mathrm{ST}}^{-1/3}([1-\xi]L_r)\nonumber\\
 &+\left[W_{\mathrm{ST}}^{2}([1-\xi]L_r)+L_\circ^2\left([1-\xi]L_r\right)/(2\pi)^2\right]^{-1/6}\Bigr\}.
\end{align}
Here  $W_{\mathrm{ST}}(L_r)$ is given by (\ref{eq:WST}) and the altitude dependence of the outer scale is given by (\ref{eq:Lout}) with $h\approx L_r\cos Z_a$.
Appendix~\ref{app:Cn2models} summarizes the model for the $C_n^2$  profile needed for the calculation of the beam wandering variance (\ref{eq:BWnew}).

\begin{figure}[ht]
 \includegraphics[width=0.45\textwidth]{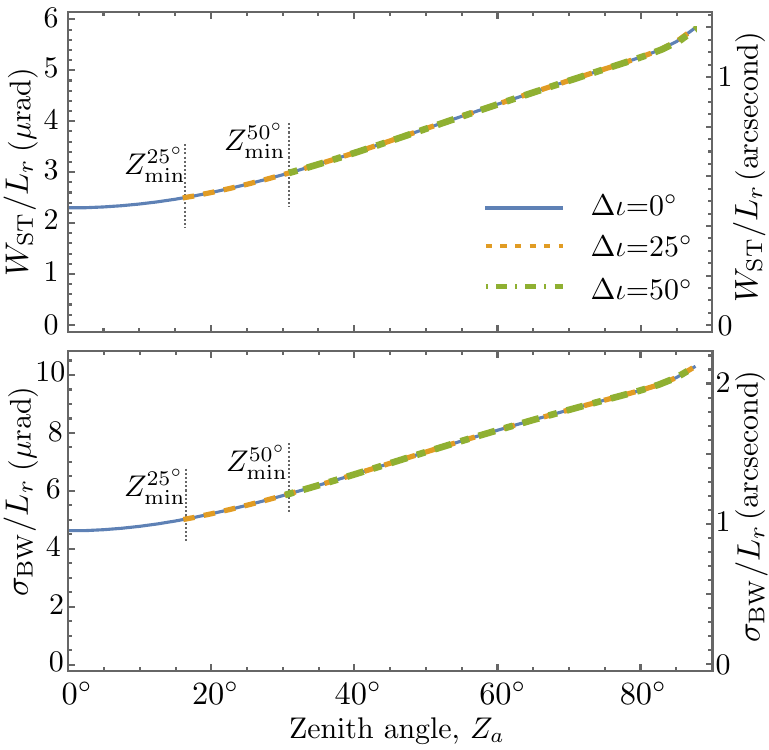}
 \caption{\label{fig:WSTBW}
Variation of the mean short-term divergence angle $W_{\mathrm{ST}}/L_r$ and the standard deviation of the angle-of-arrival  fluctuations, $\sigma_{\mathrm{BW}}/L_r$, with the zenith angle.
The results are presented for several inclination angles of the satellite orbit $\Delta\iota$.
Dotted lines indicate the minimal zenith angles (\ref{eq:Zmintext}) for inclined orbits.
}
\end{figure}

Figure~\ref{fig:WSTBW} shows  the divergence angle $W_{\mathrm{ST}}/L_r$ and the  standard deviation of the angle-of-arrival fluctuations  $\sigma_{\mathrm{BW}}/L_r$ as a function of the apparent zenith angle and several inclination angles $\Delta\iota$ of the satellite orbit.
For the simulation of turbulent atmosphere we use the AFGL+WK (the Air Force Geophysics Laboratory and Walters-Kunkel model) night model for the altitude variation of the refractive index structure parameter (see Appendix~\ref{app:Cn2models}).
Other relevant atmospheric parameters are listed in Table~\ref{tab:AtmParameters}.
The beam spot radius and the beam wandering variance grow with the growth of the zenith angle.
The asymmetry of the curves for inclined orbits is dictated by the geographical position of the observer as discussed in Sec.~\ref{sec:distance}.
For  zenith angles near $Z_a=80^\circ$ some saturation of both quantities appears.
This is due to the drop of the turbulence strength in the lower troposphere (3-10 km), the region that has its maximal contribution to the optical path at this zenith angle.
Near the observer horizon the contributions of boundary layer turbulence at 1~km as well as of the atmospheric refraction have a maximal effect on the optical beam distortions leading to the growth of both beam-spot size and beam wandering.


\section{Probability distribution of transmittance}
\label{sec:PDT}

In this section we consider the model of atmospheric quantum channels.
For the physical consistence of this model the preservation of the canonical commutation relations for the quantized optical field operators is important.
This requirement puts certain restrictions on the probability distribution that governs the statistics of fluctuating channel transmittance.
We also show the relationship of the statistical characteristics of this distribution and the moments of the field correlation functions derived in the previous section.

Quantum light transmission through a linear medium, such as Earth's atmosphere, is conveniently characterized via the  input-output relations
\begin{align}
\label{eq:inout}
 \hat a_{\mathrm{out}}=\sqrt{\eta}\hat a_{\mathrm{in}}+\sqrt{1-\eta}\hat c,
\end{align}
where $\hat a_{\mathrm{in (out)}}$ is the input (output) field annihilation operator and $\hat c$ is an environmental mode operator.
The random transmittance $\eta\in[0,1]$ equals to the instantaneously transmitted normalized intensity truncated by the receiver aperture [cf. Eq.~(\ref{eq:eta})].
The transmission of the quantum state through the atmospheric link depends not only on the characteristics of the channel, but also on the parameters of the receiver aperture.
In terms of the Glauber-Sudarshan $P$ function~\cite{Glauber1963, Sudarshan1963}, the input-output relation (\ref{eq:inout}) reads~\cite{Semenov2009} as
\begin{align}
\label{eq:Pinout}
 P_{\mathrm{out}}(\alpha)=\int\limits_0^1\D\eta\,\mathcal{P}(\eta)\frac{1}{\eta}P_{\mathrm{in}}\left(\frac{\alpha}{\sqrt{\eta}}\right),
\end{align}
where we have assumed that the environmental modes are in the vacuum state.
Here,  $P_{\mathrm{in}}(\alpha)$ and $P_{\mathrm{out}}(\alpha)$ are $P$ functions of the input and output quantum fields and $\mathcal{P}(\eta)$ is the probability distribution of the transmittance (PDT).
Hence, the description of quantum-light propagation through the atmosphere   reduces merely to identifying $\mathcal{P}(\eta)$.

An important requirement for the PDT is that its  domain of definition is restricted to the interval $[0,1]$, which is the consequence of the canonical commutation relation for $\hat a_{\mathrm{out}}$ and $\hat a_{\mathrm{out}}^\dagger$.
Violation of this requirement may lead to unphysical effects which have critical influence, e.g., on security bounds of communication protocols.
The first consistent model of the PDT \cite{Vasylyev2012} considered beam wandering as the main source of fluctuating losses on the receiver.
In this case, the PDT can be derived in analytical form and is given by the log-negative Weibull distribution.
This beam wandering model was further extended to include effects due to random beam broadening and deformation of the beam profile into an elliptic form \cite{Vasylyev2016}.
In Ref.~\cite{Vasylyev2017} this so-called elliptic-beam model was extended to include additional beam broadening and extinction due to random scattering on atmospheric aerosols and dust particles.
The elliptic-beam model gives reasonable agreement with the PDT measurements for short-distance links.
For long-distance channels, a discrepancy arises between the transmittance moments calculated via the elliptic-beam PDT and the corresponding moments (\ref{eq:eta1}) and (\ref{eq:eta2}) calculated from the first principles.
For elimination of this discrepancy, a PDT model has been proposed based on the law of total probability~\cite{Vasylyev2018}.
This model is most suitable for the description of long-length quantum channels and in this section we extend it to the case of slant propagation paths.

Analyzing experimental PDTs for short and long propagation distances, one observes two limiting behaviors.
For short distances, the beam wandering appears to be the major source of fluctuating losses and the corresponding PDT has similar form to the log-negative Weibull distribution.
For long propagation distances, beam broadening and deformation are the dominating effects yielding the PDT in the form of truncated log-normal distribution.
In the general case, using the law of total probability~\cite{Schervish1995}, the PDT can be written as
\begin{align}
\label{eq:PDTmodel}
 \mathcal{P}(\eta)=\int\limits_{\mathbb{R}^2}\D^2\boldsymbol{r}_0 P(\eta|\boldsymbol{r}_0)\rho(\boldsymbol{r_0}),
\end{align}
where the random vector $\boldsymbol{r}_0$  transverse to the propagation direction describes the position of the deflected beam centroid relative to the aperture center.
The corresponding probability distribution in (\ref{eq:PDTmodel}),
\begin{align}
\label{eq:rhoBW}
 \rho(\boldsymbol{r}_0)=\frac{1}{2\pi\sigma_{\mathrm BW}^2}\exp\left[-\frac{|\boldsymbol{r}_0|^2}{2\sigma_{\mathrm{BW}}^2}\right],
\end{align}
describes the beam-wandering contribution to the total PDT.
Here,  $\sigma_{\mathrm{BW}}^2$  is the beam wandering variance,  Eq.~(\ref{eq:BW}).
The effects of beam-spot distortions are incorporated in the conditional distribution $P(\eta|\boldsymbol{r}_0)$.
Physically it can be interpreted as the conditional PDT for the beam with a centroid position tracked to the position $\boldsymbol{r}_0$ relative to the aperture center.

For negligible small beam wandering, $\sigma_\mathrm{BW}\rightarrow 0$, the distribution (\ref{eq:rhoBW}) reduces to Dirac delta function and the conditional probability according to (\ref{eq:PDTmodel}) reduces to the PDT.
In this limit, the PDT resembles the log-normal distribution.
Hence, for general situations we can assume that  the conditional distribution can be  approximated by the truncated log-normal distribution
\begin{align}
\label{eq:CondDistr}
 &P(\eta|\boldsymbol{r}_0)\\
 &\approx\Biggl\{ \begin{array}{cc}
                                         \frac{1}{\mathcal{F}(1)}\frac{1}{\sqrt{2\pi}\eta\sigma_{r_0}}\exp\left[-\frac{(\log\eta{+}\mu_{r_0})^2}{2\sigma_{r_0}^2}\right]&\text{for}\quad \eta\in[0,1],\\
                                         0&\text{otherwise,}
                                        \end{array}\nonumber
\end{align}
where $\mathcal{F}(1)$ is the cumulative function of the (untruncated) log-normal distribution at $\eta=1$.
The parameters of this distribution are related to the conditional moments $\langle\eta\rangle_{r_0}$ and $\langle\eta^2\rangle_{r_0}$.
These conditional moments can be considered as the corresponding moments of the aperture transmittance of the effective beam with beam-spot radius $W_{\mathrm{ST}}$ whose centroid is displaced to the distance $r_0$ from the aperture center.
In the limit of weak beam wandering, the conditional moments can be written as (see Ref.~\cite{Vasylyev2018} for details)
\begin{align}
\label{eq:etacondit}
 \langle\eta\rangle_{r_0}\approx\eta_0\exp\left[-\Bigl(\frac{|\boldsymbol{r}_0|}{R}\Bigr)^\lambda\right],
\end{align}
\begin{align}
\label{eq:etacondit2}
 \langle\eta^2\rangle_{r_0}\approx\zeta_0^2\exp\left[-2\Bigl(\frac{|\boldsymbol{r}_0|}{R}\Bigr)^\lambda\right],
\end{align}
where
\begin{align}
\label{eq:eta0}
 \eta_0=\frac{\langle\eta\rangle}{\int_0^\infty\D\xi \,\xi \,e^{-\xi^2/2}e^{-[(\sigma_{\mathrm BW}/R)\xi]^\lambda}},
\end{align}
\begin{align}
 \zeta_0^2=\frac{\langle\eta^2\rangle}{\int_0^\infty\D\xi\, \xi\, e^{-\xi^2/2}e^{-2[(\sigma_{\mathrm BW}/R)\xi]^\lambda}},
\end{align}
\begin{align}
 R=a\Biggl\{\ln\biggl[2\frac{1-\exp\left(-2\frac{a^2}{W_{\mathrm{ST}}^2}\right)}{1-\exp\left[-4\frac{a^2}{W_{\mathrm{ ST}}^2}\right]\I_0\left(4\frac{a^2}{W_{\mathrm{ST}}^2}\right)}\biggr]\Biggr\}^{-1/\lambda},
\end{align}
\begin{align}
\label{eq:zeta0}
 \lambda&=8\frac{a^2}{W_{\mathrm{ST}}^2}\frac{e^{-4(a^2/W_{\mathrm{ ST}}^2)}\I_1\left(4\frac{a^2}{W_{\mathrm {ST}}^2}\right)}{1-\exp\left[-4\frac{a^2}{W_{\mathrm {ST}}^2}\right]\I_0\left(4\frac{a^2}{W_{\mathrm{ST}}^2}\right)}\\
 &\qquad\times \Biggl[\ln\biggl(2\frac{1-\exp\left(-2\frac{a^2}{W_{\mathrm{ST}}^2}\right)}{1-\exp\left[-4\frac{a^2}{W_{\mathrm {ST}}^2}\right]\I_0\left(4\frac{a^2}{W_{\mathrm{ST}}^2}\right)}\biggr)\Biggr]^{-1}\nonumber.
\end{align}
Here, $W_{\mathrm {ST}}$ is given by (\ref{eq:WST1}) and  $\I_n(x)$ is the modified Bessel function of $n$th  order.
The parameters of the conditional distribution (\ref{eq:CondDistr}) are determined from Eqs.~(\ref{eq:etacondit}) and ~(\ref{eq:etacondit2}) approximately:
\begin{align}
 \mu_r\approx-\ln\left[\frac{\langle\eta\rangle^2_{r_0}}{\sqrt{\langle\eta^2\rangle_{r_0}}}\right]\approx -\ln\left[\frac{\eta_0^2}{\zeta_0}\right]+\left(\frac{|\boldsymbol{r}_0|}{R}\right)^\lambda,
\end{align}
\begin{align}
 \sigma_{r_0}^2\approx\sqrt{\ln\left[\frac{\langle\eta^2\rangle_{r_0}}{\langle\eta\rangle^2_{r_0}}\right]}\approx\ln\left[\frac{\zeta_0^2}{\eta_0^2}\right].
\end{align}
The knowledge of the parameter set $\{\langle\eta\rangle,$ $\langle\eta^2\rangle,$ $\sigma_{\mathrm{BW}},$ $W_{\mathrm{ST}}\}$ is therefore sufficient for the determination of the channel PDT.
For detection with a Cassegrain-type telescope, the  PDT derivation procedure can be further generalized as described in Ref.~\cite{Vasylyev2018a}.

Practical optical communication via satellites is impossible without acquiring and tracking the received signal.
Both beam wandering due to atmospheric turbulence~\cite{Andrews2005} or satellite vibrations~\cite{Arnon1998} cause changes in the direction of the received beam that result in misalignment between the communication parties.
Moreover,  the velocity aberration point ahead and the atmospheric dispersion effects~\cite{Oi2017} should be taken into account.
The compensation of these disturbances requires an active beam steering that is accomplished by mechanical means.
Quantum key exchange with low mean intensities requires especially precise beam tracking and stable pointing~\cite{Moll2012, Bourgoin2015}.
Applying coarse and fine tracking and pointing strategies, one can achieve a tracking accuracy of approximately $\theta_{\mathrm{tr}}\sim 1.2 \mu$rad for LEO-ground communication links~\cite{Liao2017a}.
We incorporate the tracking procedure in the PDT model (\ref{eq:PDTmodel}) by replacing
\begin{align}
\label{eq:Btracking}
 \sigma_{\mathrm{BW}}\rightarrow\sigma_{\mathrm{tr}}=\theta_{\mathrm{tr}}L_r(Z_a)
\end{align}
in Eq.~(\ref{eq:rhoBW}).
Finally, we note that the parameters (\ref{eq:eta0}) and (\ref{eq:zeta0}) still contain the beam wandering variance, and we calculate $\sigma_{\mathrm{BW}}^2$ by means of Eq.~(\ref{eq:BWnew}).


\section{Application: Decoy state protocol}
\label{sec:Decoy}

We apply the developed theory of satellite-mediated quantum atmospheric channels for the calculation of the secret key rate when decoy states are in use~\cite{Hwang2003, Lo2005}.
Conventionally, we refer to communication parties  Alice and Bob as to the sender and the receiver, respectively.
Based on the BB84 protocol~\cite{Bennett1984}, the decoy-state method estimates channel parameters by sending two types of states.
While one type of states (signal states) is used for transmission of quantum keys, the other is called the decoy state, which is used for the estimation of the number of transmitted  single-photon states.
Ideally, the single-photon states are most suitable to be used as the signal.
Practically perfect single-photon sources are hard to attain and one uses weak coherent states instead.
In the security analysis of the decoy-state method, both signal and decoy states possess equal properties except for their intensity.

Usually, only a few decoy states are needed for practical implementations.
A simple two-decoy-state protocol with vacuum+weak decoy states gives an optimal key generation rate which is the same as having an infinite number of decoy states~\cite{Ma2005}.
On the first stage of the protocol, Alice, who has a phase-randomized source of coherent states, encodes the bits in the $X$ or $Z$ basis as in the standard BB84 scheme, e.g., by utilizing polarization degrees of freedom.
Additionally to the signal field, she generates decoy states in vacuum and weak coherent states.
The phase randomization makes the source statistically equivalent to a Poissonian distribution of Fock states such that, when the average photon number from the light source is $\mu$, the probability to send an $n$-photon pulse is $e^{-\mu}\mu^n/n!$.
We denote the mean photon numbers as $\mu_j$, $j=s,d,v$ for signal, weak-decoy, and vacuum states.
The following conditions are satisfied: $\mu_d<\mu_s<1$, $\mu_v=0$.
After transmission through the free-space channel, Bob performs measurement of transmitted bits in a randomly chosen  $X$ or $Z$ basis.
The conditional probability of a detection event at Bob's side given that Alice sends an $i$-input state is referred to as the yield $Y_i$ of an $i$-photon state.
The vacuum state is used for the estimation of the background detection probability $Y_0$ while the weak-decoy state allows one to estimate the single-photon yield $Y_1$ and the error rate of the single-photon state $e_1$.
Since both signal and weak-decoy states propagate through the same channel, the single-photon yield would be the same for these states.
Due to this property, the security of the decoy-state protocol against the photon-number splitting attacks~\cite{Scarani2009, Pfister2016} or the Trojan-horse attacks~\cite{Tamaki2016} can be verified.

As the next step, the parties perform the sifting of the raw key, its error correction, and privacy amplification.
Finally, if these steps were successful, Alice and Bob share a shorter but more secure key.
Defining the  averaging over channel fluctuations  with the applied beam tracking procedure  (\ref{eq:Btracking}) as
\begin{align}
\label{eq:etatr}
 \langle f(\eta)\rangle_{\mathrm{tr}} =\int\limits_0^1\D\eta\, f(\eta)\,\mathcal{P}(\eta)\Bigr|_{\sigma_{\mathrm{BW}}\rightarrow\sigma_{\mathrm{tr}}},
\end{align}
we find for the lower bound of the average secure key rate
\begin{align}
\label{eq:keyrate}
 \mathcal{R}&= q \Bigl(-\langle Q_{\mu_s}\rangle_{\mathrm{tr}}f(\mathrm{QBER}) H[\mathrm{QBER}]\\
 &+\sum\limits_{\gamma=x,z}\langle Q_1^{\gamma L}\rangle_{\mathrm{tr}}\{1-H[\langle e_1^{\gamma U}Q_1^{\gamma L}\rangle_{\mathrm{tr}}/\langle Q_1^{\gamma L}\rangle_{\mathrm{tr}}]\} \Bigr).\nonumber
\end{align}
Here, $q$ depends on the implemented protocol, $H(x)=-x\log_2(x)-(1-x)\log_2(1-x)$ is the binary Shannon information function, $Q_{\mu_s}$ is the gain of the signal states,  $Q_1$ is the gain of single-photon states, and $f(x)\ge1$ is the bidirectional error correction efficiency ($f(x)=1$ corresponds to the perfect error correction case).
The quantum bit error rate, QBER, is estimated as
\begin{align}
\mathrm{QBER}=\frac{\langle E_{\mu_s}Q_{\mu_s}\rangle_{\mathrm{tr}}}{\langle Q_{\mu_s}\rangle_{\mathrm{tr}}}.
\end{align}
The gain $Q_{\mu_s}$ represents the ratio between the number of events where Bob observes a click under the condition that Alice sends a certain number of signal states.
This overall gain with respect to an ideal threshold detector~\cite{Kok2007} can be evaluated as~\cite{Ma2005}
\begin{align}
\label{eq:OverallGain}
 Q_{\mu_s}=\sum\limits_{i=0}^\infty Q_i^s=\sum\limits_{i=0}^\infty Y_i\frac{\mu_s^i}{i!}e^{-\mu_s}=1-e^{-\eta_{\mathrm{d}}\eta\mu_s}(1-Y_0),
\end{align}
where the yield of the $i$-photon state is
\begin{align}
 Y_i= 1-(1-Y_0)(1-\eta_{\mathrm{d}}\eta)^i.
\end{align}
Here
\begin{align}
\label{eq:etad}
\eta_{\mathrm{d}}=\eta_{\mathrm{det}}\chi_{\mathrm{ext}}\chi_{\mathrm{opt}}
\end{align}
accounts for deterministic losses such as detector efficiency, channel losses due to atmospheric extinction (\ref{eq:chiExt}), and absorption by optical components, while $\eta$ is the random transmittance of the free-space channel [cf. Eq.~(\ref{eq:eta})].
Similarly, defining the error rate of the $i$-state as
\begin{align}
 e_iY_i=e_0 Y_0+e_{\mathrm {det}}[1-(1-\eta_{\mathrm{d}}\eta)^i](1-Y_0),
\end{align}
one derives for the overall error gain
\begin{align}
\label{eq:OverallQBER}
 E_{\mu_s} Q_{\mu_s}&=\sum\limits_{i=0}^\infty e_i Y_i\frac{\mu_s^i}{i!}e^{-\mu_s}\\
 &=e_0 Y_0+e_{\mathrm{det}}(1-e^{-\eta_{\mathrm{d}}\eta\mu_s})(1-Y_0).\nonumber
\end{align}
Here, $e_{\mathrm{det}}$ is the probability that an incorrect bit value occurred  that depends on the alignment and the stability of the optical system.
The background  error rate is $e_0=1/2$ for randomly occurring dark and background counts.
The dark count contribution to $Y_0$ is of order $10^{-6}$ for a commercially available Geiger-mode APD at room temperature and can be further decreased with proper cooling.
The contribution from transmitted vacuum decoy states with the accounting of finite-size effects is given in Appendix~\ref{app:Decoy}.
The total value $Y_0$ can be further enhanced due to sky-noise photodetection~\cite{Erlong2005, Gruneisen2015, Nordholt2002}.
In this article we assume that the background error rate is constant with the value taken from Ref.~\cite{Liao2017}.

For the determination of the single-photon gain, $Q_1$, and the single-photon error rate, $e_1$, in Eq.~(\ref{eq:keyrate}) the statistical fluctuations must be considered.
Indeed, since the communication link with a LEO satellite can be established for only several minutes, only a finite set of data can be transmitted.
In the security analysis the accounting for possible deviations from most probable values must be taken into account.
Moreover, statistical fluctuations tend to become more important as the distance of the QKD increases, i.e. for large values of the zenith angle.
In Ref.~\cite{Ma2005} the Gaussian model and in Refs.~\cite{Curty2014, Lim2014} the Chernoff-Hoeffding method have been applied for deriving  finite-key security bounds.
We adopt the statistical fluctuation analysis of Ref.~\cite{Zhang2017}  that uses the Chernoff bound for establishing the lower bound  for single-photon gain,
\begin{align}
Q_1^{\gamma L}(\eta)=Y_1^{\gamma L}(\eta)\mu_s e^{-\mu_s},\quad \gamma=x,z,
\end{align}
and the relation between the upper bounds for  single-photon error rates,
\begin{align}
e_1^{zU}(\eta) = e_1^{xU}(\eta)+\theta^U,
\end{align}
in $X$ and $Z$ bases.
Appendix~\ref{app:Decoy} summarizes the method for obtaining $\theta^U$ [cf. Eq.~(\ref{eq:epsilonTheta})], the lower bound for the single-photon yield $Y_1^{\gamma L}$ [cf.~Eq~(\ref{eq:yield})] and the upper bound for the bit-flip error rate $e_1^{xU}$ [cf.~Eq.~(\ref{eq:errorxU})].

We calculate the secure key rate for the downlink communication scenario.
Table \ref{tab:DecoyParameters} lists the values of parameters used in the calculation, whereas Table~\ref{tab:AtmParameters} gives parameters of the atmospheric link and communication system.
With the source of repetition rate 150 MHz, the total number of generated bits by Alice during 14 minutes of communication session is $N=10^{11}$.
We estimate the mean number of sifted key bits,
\begin{align}
 M^a\approx\eta_{\mathrm{sift}}\eta_{\mathrm{d}}\eta N^a,\quad a=s,d,v,
\end{align}
where $\eta_{\mathrm{sift}}=0.5$ is the sifting efficiency for conventional sifting protocols.
The number of sifted key bits is averaged over the fluctuations of the channel transmittance and depends on the zenith angle (e.g., for zenith orbit  $\langle M^s\rangle_{\mathrm{tr}}=1.46{\times}10^{5}$, $\langle M^d\rangle_{\mathrm{tr}}=5.64{\times}10^{4}$ for $Z_a=0^\circ$,  and $\langle M^s\rangle_{\mathrm{tr}}=1.93{\times}10^{4}$, $\langle M^d\rangle_{\mathrm{tr}}=7.43{\times}10^{3}$ for $Z_a=70^\circ$).

\begin{table}[h]
\caption{\label{tab:DecoyParameters}
Parameters for a QKD system.
}
\begin{ruledtabular}
\begin{tabular}{ l c c }
\textrm{Parameter}&\textrm{Notation}&\textrm{Value}\\
\colrule
Detector efficiency       & $\eta_{\mathrm{det}}$     &  $60\%$             \\
Extinction due to optics     & $\chi_{\mathrm{opt}}$     &  $84\%$               \\
Background yield (dark count)        & $Y_0^{\mathrm{DC}}$                     &  $5.89\times10^{-7}$  \\
Erroneous detector prob.  & $e_{\mathrm{det}}$        &  $1\%$               \\
Background error rate     & $e_0$                       & $50\%$  \\
Failure probability       & $\varepsilon$             &  $10^{-5}$        \\
Mean intensity of signal  & $\mu_s$                   &  $0.8$  \\
\qquad weak decoy state   & $\mu_d$                   &  $0.1$  \\
Number of sent bits       & $N$                       &  $10^{11}$             \\
Pulse repetition rate     & $r_N$  (MHz)                   &  $150$             \\
Generation prob. signal bits & $p_s$                  &  $65\%$             \\
\qquad weak decoy bits    & $p_d$                     &  $25\%$             \\
Generation prob. of x-basis bits   & $p^x_a$, $(a{=}s,d)$                     &  $60\%$             \\
Error correction efficiency   & $f(\text{QBER}) $                     & $1.16$             \\
Tracking precision       & $\sigma_{\mathrm{tr}}$ ($\mu$rad)     & $1$            \\
\end{tabular}
\end{ruledtabular}
\end{table}

\begin{figure}[ht]
 \includegraphics[width=0.46\textwidth]{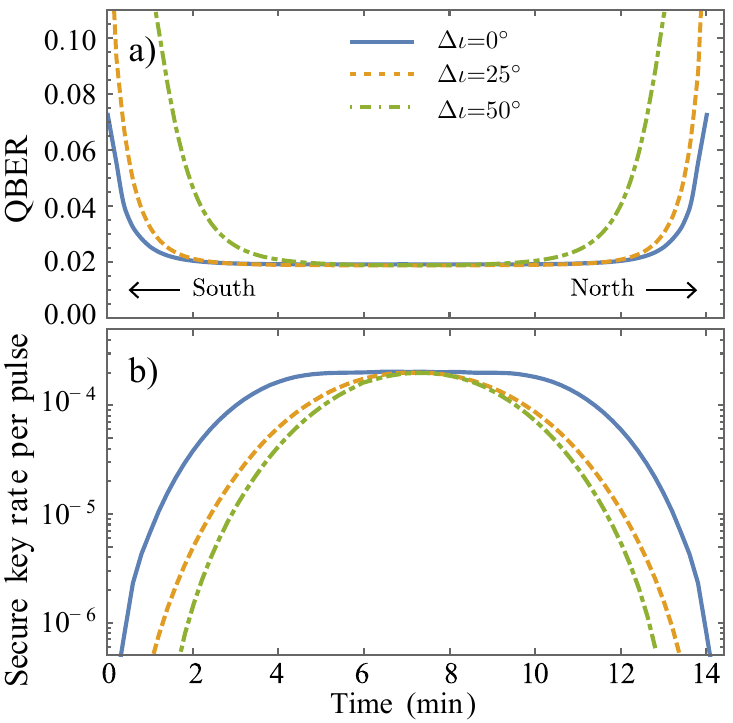}
 \caption{\label{fig:KeyRate}
Quantum bit error rates (a) and  lower bounds of averaged secure key rates (b) as  functions of the communication time for several inclination angles.}
\end{figure}

Figure~\ref{fig:KeyRate} shows the QBER and the lower bound of the secure key rate calculated for several inclination angles of satellite orbits relative to the observer's meridian plane.
For small zenith angles scintillations, atmospheric refraction, and absorption  have minimal impact on the performance of  decoy-state protocols.
The QBER has almost constant value, the phenomenon that was theoretically predicted for satellite links in Ref.~\cite{Shapiro2011} and experimentally observed in Refs.~\cite{Casado2017, Liao2017, Liao2017a}.
Figure~\ref{fig:KeyRate} a) shows a similar behavior for the zenith angles where the aperture-averaged scintillation index has the asymptotic behavior (\ref{eq:scintEta}) and satisfies the condition $\sigma^2_\eta<1$.
In the region of saturated scintillations the QBER grows making quantum communication impossible for large zenith angles.
In this region the wavefront distortions are maximal.
These distortions together with the extinction losses lower the signal-to-noise ratio and correspondingly the QBER is inflated by the larger relative contributions from the background.
The flat region of the QBER diminishes with the growing inclination angle of the satellite orbit allowing smaller time windows for the secure key exchange as shown in Fig.~\ref{fig:KeyRate} b).

It is worth to note that by placing the observer at high altitudes above sea level this time window can be increased.
Indeed, the thickness of the dense ground layer, which is responsible for scintillation, is different at sea level and at high altitudes.
This causes a shift of the region with saturated scintillations to higher zenith angles.
This phenomenon can be observed if one compares the curves  of Fig.~\ref{fig:SciRigorous} calculated for the observer at sea level with the experimental curve of Fig.~\ref{fig:Scintillation} with  observer's elevation of 602 m above sea level.
The saturation region may even vanish at particularly high altitudes, the phenomenon which is  known in optical astronomy~\cite{Kucherov, Fuentes1987, Butler1952}.
This observation makes observatories in mountains especially attractive as a OGS node for quantum free-space communication with satellites.


\section{Conclusions}
\label{sec:Conclusions}

In this article we have presented the theoretical analysis of satellite-mediated quantum links.
We have discussed the influence of regular  refraction, extinction, and turbulence on the transmission properties of optical signals through the Earth's atmosphere.
Since in satellite-mediated communication scenario the position of the satellite changes rapidly for the observer located on the Earth's surface, the thoughtful analysis is presented of how atmospheric disturbances depend on the observer's geographical position, observer's zenith angle, and  the orbit inclination angle.

We focus our analysis on low orbit satellites with perfect polar orbit.
In particular, we considered the case when the satellite orbit is inclined to the observer meridian and derived the corresponding slant range.
The orbit inclination restricts the definition range of the zenith angle by introducing the lower bound of the angle.
As a consequence, the most favorable satellite trajectory corresponds to the zenith orbit that passes through the observer meridian.
In this case, the smallest zenith angle is zero and the slant range is the shortest one at the observer zenith.

The effect of regular atmospheric refraction increases the optical slant range and changes the value of the true zenith angle to the apparent one.
This effect is especially pronounced near the horizon where it can increase the propagation path of the optical signal up to $30\%$.
Based on the standard atmosphere model we derived the corresponding path elongation factor as a function of the apparent zenith angle  and gave the corresponding analytical fit formula.
The resulting slant range has been used for the calculation of the atmospheric extinction factor due to absorption and scattering on atmospheric gases and aerosols.

Another important factor that deteriorates the optical performance of satellite-ground links is the atmospheric turbulence.
It appears that the second- and the fourth-order optical field correlation functions play a central role in the description of light propagation through the turbulent media.
These functions allow one to derive the aperture-averaged scintillation index, as well as the mean beam-spot radius and the beam wandering variance of the transmitted beam.
Based on the properties of intensity covariance function, we have derived an analytical expression for the aperture-averaged scintillation index.
For large zenith angles these scintillations  saturate and decrease; this effect  is well observed in experiment.
It arises due to multiple scattering in most turbulent air layers near the ground, the process that degrades the performance of the receiver telescope and leads to the additional aperture-averaging of scintillations.
Based on this simple model of aperture averaged scintillations, we have developed the rigorous approach for calculating field correlation functions and their moments.

The developed description of atmospheric channels has been adopted for the description of quantum light propagation through the Earth's atmosphere.
For this sake we use the input-output relations for optical field operators, rewritten in terms of Glauber-Sudarshan $P$ function.
Fluctuations of the channel transmittance due to atmospheric turbulence are accounted with the help of the probability distribution of transmittance.
We have obtained the latter for satellite-mediated quantum links using the law of total probability.

Finally, the security of quantum decoy-state communication  protocols is analyzed with realistic channel parameters and communication conditions.
In this connection, the inclusion of the finite-key effects plays an important role.
Scintillation phenomena at large zenith angles influence greatly the performance of quantum channels, leading to a growth of the quantum bit error rate.
In the region of saturated scintillations that appears close to the horizon, no secure quantum key can be obtained.
For small zenith angles the quantum bit error rate has minor variation that allows one to obtain secure key bits within a certain time window.
This window increases  with the decrease of the satellite inclination angle relative to observer's meridian plane.
We also expect that the increase of the observer's altitude relative to the sea level will lead to a decrease of the region with saturated scintillations and hence to a better communication performance.

We have omitted several  aspects affecting the performance of the satellite-mediated quantum communication.
Background radiation  from the Sun, Moon, stars, or light reflected from the satellite introduce additional noise.
This noise is detected by photodetectors, and therefore, it increases the quantum bit error rate.
The use of light buffers, time gate, frequency, and spatial filters can partially mitigate the problem of background noise.
In this study we have neglected the parallax-connected errors due to the Earth's rotation during one communication session.
For observers at small geographical latitudes, and for the satellites with small altitude such parallax effects should be included in the rigorous analysis.
On the other hand, for the analysis of realistic communication scenarios not only the instantaneous position of a satellite relative to the observer is important, but also the relative position of the Sun or of the Moon.
We refer to extensive literature that study the influence of background noise and dark counts on the satellite-mediated quantum key distribution.

Much less studied is the influence of random scattering on atmospheric aerosols and dust particles on the performance of satellite-based communication.
Especially in overcast conditions the dominance of Mie scattering makes the optical quantum communication with satellites impossible.
We have included in the model atmospheric extinction effects due to absorption and scattering on molecules and aerosols which correspond to average communication conditions.
Under the conditions of low visibility and overcast skies, the performed analysis is not applicable.

Finally, here we have considered fundamental spatial modes for the light beams that have Gaussian profiles of the intensity distributions.
The efficient generation of   diffraction-free beams will allow one to reduce losses associated with regular and random diffraction in free-space and may extend the range of satellite-mediated quantum communication.

\acknowledgments The work was supported by the Deutsche Forschungsgemeinschaft through projects VO 501/21-2 and VO 501/22-2.
The authors are grateful to A. A. Semenov for useful and enlightening discussions.

\appendix

\section{Slant range of optical beam}
\label{app:distance}

In this appendix we derive the length of geometrical distance between the satellite and the optical ground station.

\subsection{Geometry of satellite communication links}

The geometric path length   for a specific communication scenario is determined from the value of the zenith angle $Z$, which in turn is a function of the satellite orbit inclination, the instant position of the satellite, and geometric latitude of the observer.
In the Earth surface coordinate system of the observer the position of the satellite is given by the zenith angle $Z$ and the azimuth angle $A$.
The latter we measure from the north point eastwards.
In the Earth center (geocentric) coordinate system the satellite position is determined by its declination angle  and its orbit inclination angle.
For circular orbits the declination angle is given by
\begin{align}
 \delta=\omega_{\mathrm{sat}}(t{-}t_{\mathrm{Pole}}),
\end{align}
where $\omega_{\mathrm{sat}}=2\pi/T_{\mathrm{sat}}$ is the angular speed of the satellite with orbiting period $T_{\mathrm{sat}}$ and $t_{\mathrm{Pole}}$ is the reference time associated with the trajectory crossing either the North or South Pole.
During a single communication session the declination angle changes from $\delta_0$ to $\delta_0+\delta_{\mathrm{com}}$ with $\delta_0$ being the initial declination of the satellite when it appears on the observer horizon and
\begin{align}
 \delta_{\mathrm{com}}=\omega_{\mathrm{sat}}t_{\mathrm{com}}=2\arctan\Bigl[\frac{\sqrt{H^2+2R_\oplus H}}{R_\oplus}\Bigr].
\end{align}
Here $t_{\mathrm{com}}$ is the maximal duration of the communication session,  $H$ is the satellite altitude above the ground, and $R_\oplus$ is the Earth radius.
Finally, the inclination angle $\Delta\iota$ we define as the angle between the observer meridian plane and the satellite orbit plane and the geographical latitude of the observer we denote as $\Psi$.
The set $\{\Psi,\delta,\Delta\iota\}$ determines the instantaneous position of the satellite relative to the observer in the geocentric coordinate system and is the alternative parameter set to the set $\{A,Z\}$ in the Earth surface coordinate system.

\begin{figure}[ht]
 \includegraphics[width=0.45\textwidth]{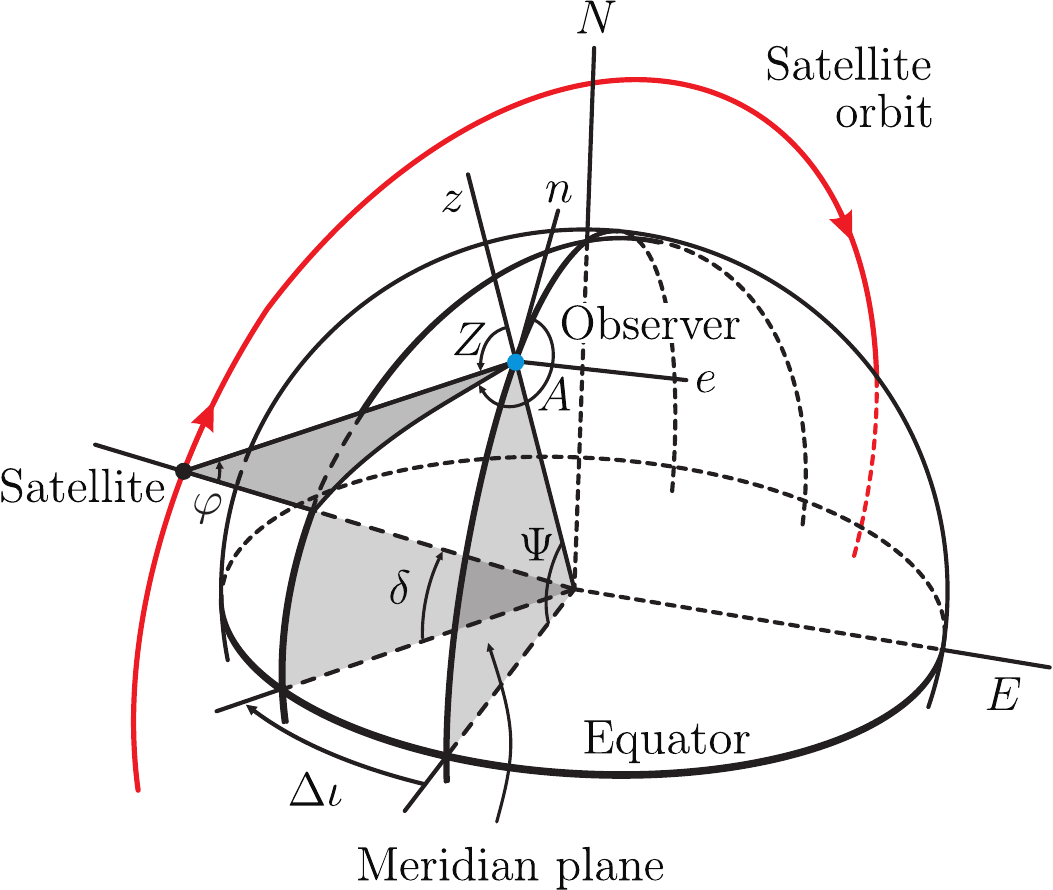}
 \caption{\label{fig:Geocentrical}
Geometric representation of the surface and geocentric coordinate systems.
The Earth surface coordinate system for the  observer showing the azimuth $A$ and the zenith $Z$ angles.
The geocentric coordinate system is defined in terms of the declination angle $\delta$ and the inclination angle $\Delta\iota$.
The geographical position of the observer is given by the latitude $\Psi$ while the parallax correction caused by the finiteness of the satellite altitude $H$ is determined by the angle $\varphi$.
}
\end{figure}

The following relation between angles $Z$,  $\delta$, $\Delta\iota$, $\Psi$ shown in Fig.~\ref{fig:Geocentrical} can be established using spherical trigonometry
\begin{align}
\label{eq:cosz1}
 \cos \xi=\sin \Psi\, \sin\delta+\cos \Psi\, \cos\delta\,\cos\Delta\iota ,
\end{align}
\begin{align}
\label{eq:cosz}
 Z=\arctan\left(\frac{\sin\xi}{\cos\xi-\frac{R_\oplus}{R_\oplus+H}}\right).
\end{align}
Here $\delta$ changes during the communication session from $\delta_0=-\arctan(\cos\Delta\iota\cot\Psi)$ to $\delta_0+\delta_{\mathrm{com}}$.
We note that the finite value of the inclination angle $\Delta\iota$ in (\ref{eq:cosz}) restricts the definition domain of $Z$ to $[Z_{\mathrm{min}}^{\Delta\iota},\pi/2]$ for the observer located at the geographical latitude $\Psi$.
The minimal value of $Z$ can be found from Eq.~(\ref{eq:cosz}) as
\begin{align}
\label{eq:Zmin}
 Z_{\mathrm{min}}^{\Delta\iota}= \arccos\left[\sqrt{1-\cos^2\Psi\sin^2\Delta\iota}\right],
\end{align}
which corresponds to the satellite declination angle (counted from the Equator)
\begin{align}
\delta_{\mathrm{min}}^{\Delta\iota}=\arccos\left[\frac{\cos\Psi\cos\Delta\iota}{\sqrt{1-\cos^2\Psi\sin^2\Delta\iota}}\right].
\end{align}
The knowledge of the satellite declination angle, the orbit inclination angle, and geographical position of the observer allows one to determine the instantaneous values of zenith angle from Eqs.~(\ref{eq:cosz}) as well as to determine the slant range.

\subsection{Inclined and zenith orbits}

\begin{figure}[ht]
 \includegraphics[width=0.35\textwidth]{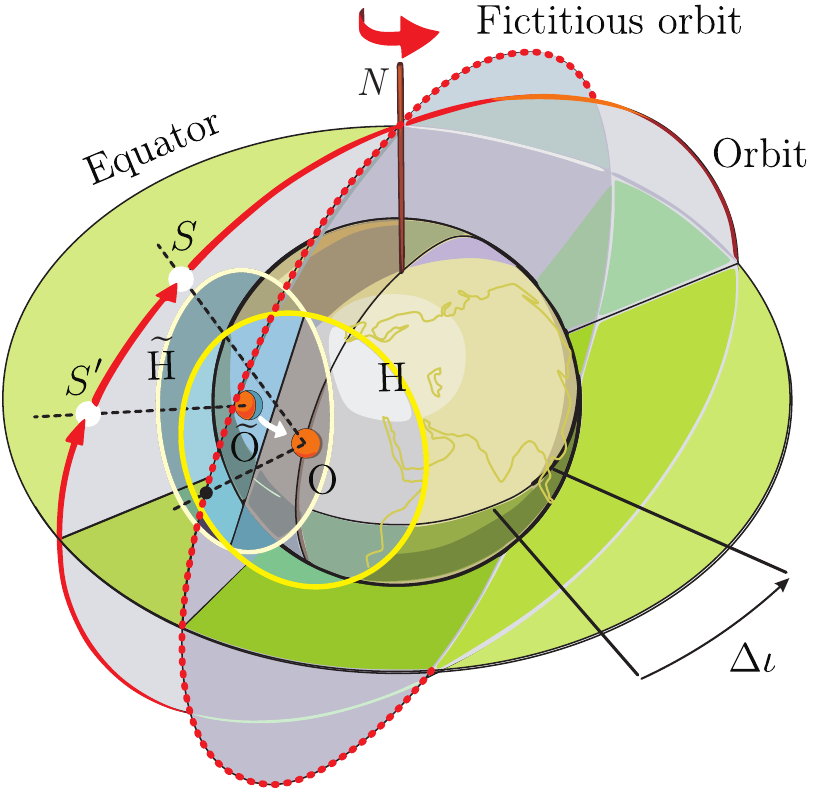}
 \caption{\label{fig:PolarNew}
 Geometry of a typical satellite-ground communication.
 Initially, the ground observer $\widetilde O$ establishes the communication link with the satellite whose trajectory lies within the meridian plane of the observer $\widetilde O$.
 At a later time, due to Earth's rotation, the observer is translated to the point $O$ while the satellite trajectory is inclined by an angle $\Delta\iota$ relative to the new observer meridian plane (or fictitious orbit plane).
 Circles $\widetilde{\mathrm{H}}$ and $\mathrm{H}$ denote the local horizons of the observers $\widetilde O$ and $O$, respectively.
 The communication between the observer $O$ and the satellite $S$ can be established provided the satellite trajectory $S^\prime S$ lies above the horizon $\mathrm{H}$.
}
\end{figure}

In most practical cases the satellite trajectory is inclined relative to the observer zenith direction.
For example, the initial trajectory with zero inclination to the observer's zenith becomes inclined when the satellite makes one or more revolutions.
Let us consider the satellite  whose orbit initially  has zero inclination relative to the meridian plane of the observer $\widetilde O$ (see Fig.~\ref{fig:PolarNew}).
Let us also assume that initially the satellite is positioned at observer's zenith, i.e., at point $S^\prime$.
For simplicity we consider the ideal polar orbit that passes through both Earth's celestial poles.
After time $T_{\mathrm{sat}}$, i.e., the satellite orbiting period, the satellite is positioned again in the point $S^\prime$.
Meanwhile the observer position $\widetilde O$ is moved to the point $O$ due to Earth's rotation.
The resulting inclination angle reads as
\begin{align}
\label{eq:iota}
 \Delta\iota=\frac{T_{\mathrm{sat}}v_\oplus}{ R_\oplus},
\end{align}
where $v_\oplus$ is the speed of Earth's rotation at the Equator.
Here and in the following we assume that the change of the inclination angle during the satellite transition over the observer horizon is relatively small, so that we can approximate it being constant (\ref{eq:iota}) during the whole communication session.
The negative values of $\Delta\iota$ refer to situations when the visible satellite trajectory is positioned eastwards to the observer.
We calculate the slant range between $O$ and $S$ along the satellite orbit provided the trajectory $S^\prime S$ of the moving satellite lies above the observer horizon $\mathrm{H}$.

The instantaneous position of the satellite relative to the observer is determined by zenith $Z$ and azimuth $A$ angles as shown in Fig.~\ref{fig:Geocentrical}.
For satellite tracking purposes both these angles are of importance.
In the context of optical satellite-based communication we are interested primarily in the instantaneous value of the slant range as a function of zenith angle.
Indeed, by assuming that tracking systems point correctly the telescope towards the flying satellite, we can can ignore the dependence of the slant range on the azimuth angle.
Then, the slant range $L$ for the inclined orbit can be determined from the triangle $OSC$ in Fig.~\ref{fig:Geometry1} using the law of cosines
\begin{align}
\label{eq:Pythagorean}
L^2+R_\oplus^2-2L R_\oplus\cos(\pi-Z)=(R_\oplus+H)^2.
\end{align}
Solving this equation with respect to $L$ we obtain
\begin{align}
\label{eq:slantrange}
L(Z)=\sqrt{H^2+2H R_\oplus+R_\oplus^2\cos^2 Z}-R_\oplus\cos Z.
\end{align}
The root of the quadratic equation (\ref{eq:Pythagorean}) is chosen to yield the non-negative slant range.
At the observer's horizon the slant range reaches its maximal value $ L(90^\circ)$$=\sqrt{(R_\oplus+H)^2-R_\oplus^2}\approx\sqrt{2H R_\oplus}$.
The shortest slant range is determined as $L(Z_{\mathrm{min}}^{\Delta\iota})$, where the minimal zenith angle is given by Eq.~(\ref{eq:Zmin}).

The special case of the slant range (\ref{eq:slantrange}) corresponds to the zenith satellite orbit with zero inclination angle $\Delta\iota$.
Some authors refer to this orbit as to the "best pass" \cite{Bourgoin2013} since the slant range, that we denote as $L_0$, gives the shortest distance between the observer and the satellite at zenith, namely, $L_0(0^\circ)=H$.
This fact makes the zenith orbit as well as the orbits with small inclination angle $\Delta\iota$  most attractive for the optical communication.
Denoting the zenith angle within the zenith orbit plane (observer's meridian plane) as $Z_0$, it follows from Eq.~(\ref{eq:cosz}) that $Z_0=\Psi-\delta$.
Consequently, we have $L_0(Z_0)=L(Z)\bigl|_{\Delta\iota=0}$.

\section{Standard Atmosphere}
\label{app:standardAtmosphere}

The standard atmosphere  is an idealized model of the Earth's atmosphere for heights ranging from the surface to 1000 km~\cite{StandardAtmosphere}.
The model yields the air density, viscosity, etc., for various altitudes.
For our purposes, the most useful values calculated within the standard atmosphere model are the temperature $T$ and the pressure $P$.
These variables are important  for the calculation of  the refractive index variation with altitude.

The altitude dependence of the temperature is approximated with linear segments
\begin{align}
\label{eq:temper}
 T(h)=T_b+\left(\frac{\D T}{\D h}\right)_b(h-H_b).
\end{align}
Each segment lies within an atmospheric layer bounded by the surfaces with the altitudes $H_{b-1}$ and $H_b$.
Table ~\ref{tab:StAtm} summarizes the reference values $T_b$ and the vertical gradient values of temperature (lapse rates)
\begin{align}
 \lambda_b=(\D T/\D h)_b,
\end{align}
which enter Eq.~(\ref{eq:temper}).
The altitude dependence of pressure can be found from the gas law and hydrostatic equation and reads as
\begin{align}
\label{eq:pressure}
 P=P_b\left\{1+\frac{\lambda_b}{T_b}(h-H_b)\right\}^{-g/\lambda_b R}
\end{align}
for constant lapse rate and
\begin{align}
\label{eq:pressure1}
 P=P_b\exp\left[-\left(h-H_b\right)g/R T_b\right]
\end{align}
for isothermal layers ($\lambda_b=0$).
Here $g=9.8~\text{m/s}^2$  is the gravitational acceleration and $R=287.053~\text{J/kg}\cdot \text{K}$ is the gas constant for air.
The reference values of pressure $P_b$ are given in Table~\ref{tab:StAtm}.

\begin{table}[h]
\caption{\label{tab:StAtm}%
The reference altitudes and values and gradients of the linearly segmented temperature-height and pressure-height profiles from the Earth's surface up to altitude of 85 km.
}
\begin{ruledtabular}
\begin{tabular}{ l c c c c }
\textrm{Sub-}&\textrm{Height}&
\textrm{Temperature}&\textrm{Temperature}&\textrm{Pressure}\\
\textrm{script}&&
\textrm{gradient}&\textrm{}&\\
$b$& $H_b$, (\textrm{km})&
$\lambda_b$, (K/km)& $T_b$, (K)&$P_b$, (mb)\\
\colrule
0& 0   & -6.5 & 288 &  1\,013\\
1& 11  &  0.0 & 217 &  226\\
2& 20  & +1.0 & 217 &  54.7\\
3& 32  & +2.8 & 229 &  8.68\\
4& 47  &  0.0 & 271 &  1.11\\
5& 51  & -2.8 & 271 &  0.67\\
6& 71  & -2.0 & 215 &   0.04\\
7& 84.8&   -   & 188 &  0.004\\

\end{tabular}
\end{ruledtabular}
\end{table}

Following Birch and Downs~\cite{Birch1994} we adopt the revised form of the  Edl\'en equation for the atmospheric refractive index $n$,
\begin{align}
 (n-1)&=\frac{(P/\text{Pa})(n-1)_s}{96095.43}\nonumber\\
 &\times\frac{1+10^{-8}(0.601-0.00972 T/{{}^\circ \text{C}})P/\text{Pa}}{1+0.0036610\, T/{{}^\circ \text{C}}},
\end{align}
where $(n-1)_s$ is given by the dispersion equation
\begin{align}
 (n-1)_s\times 10^8&=8342.54\nonumber\\
 &+2406147\left[130-\left(1/\lambda\right)^2\right]^{-1}\\
  &+15998\left[38.9-\left(1/\lambda\right)^2\right]^{-1}\nonumber.
\end{align}
Here $\lambda$ is the optical wavelength given in $\mu$m.
Since the values of  the refractive index are distinct from the vacuum value $n=1$ up to the mesosphere, we restrict our attention only to altitudes ranging from 0 to 85 km.

\begin{figure}[ht]
 \includegraphics[width=0.48\textwidth]{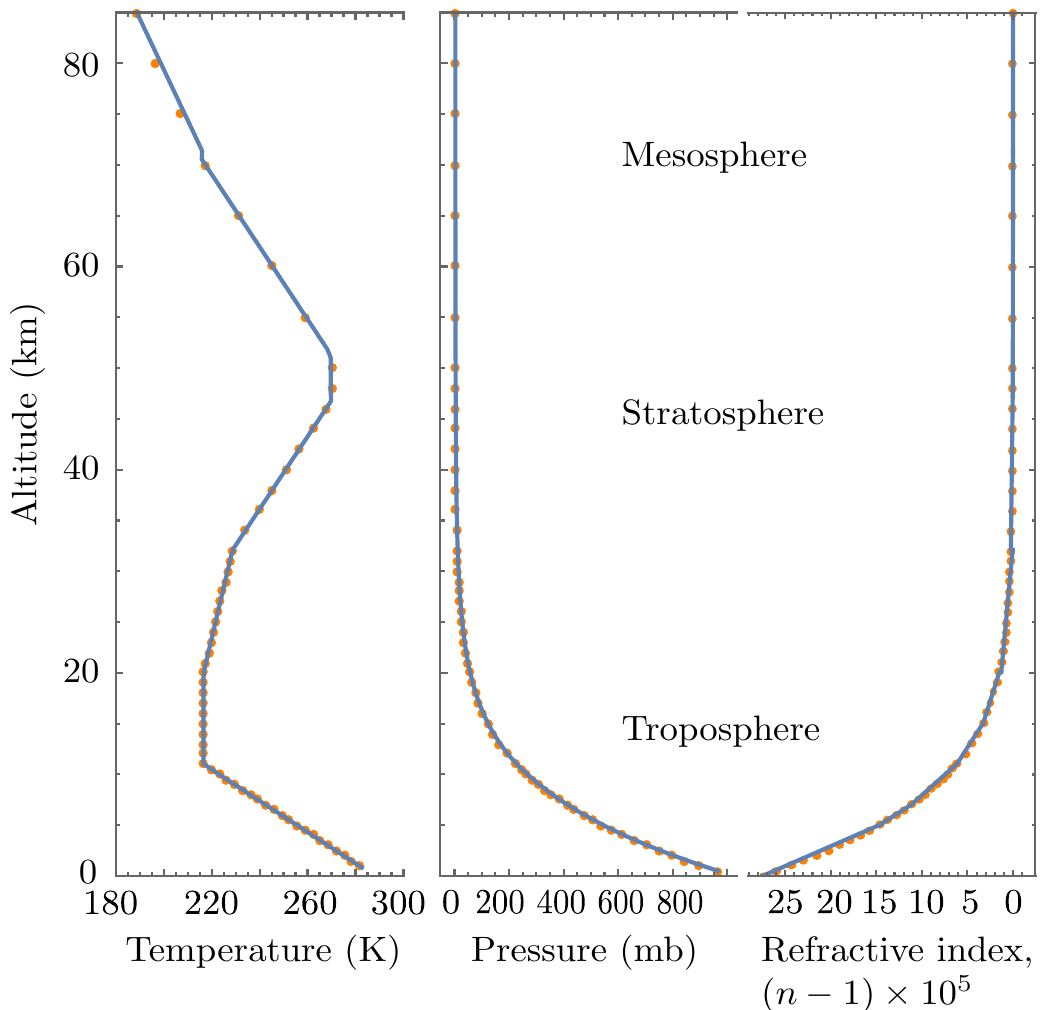}
 \caption{\label{fig:Parameters} Temperature, pressure and refractive index as  functions of altitude.
 Dots represent the standard atmosphere values and lines show the  linear approximation of temperature and refractive index curves used in this article, as well as
 the functional dependence (\ref{eq:pressure}), (\ref{eq:pressure1}) for the pressure.
}
\end{figure}

Figure \ref{fig:Parameters} shows the variation of temperature, pressure and refractive index with altitude.
The temperature profile shows several layers where temperature dependence on altitude  can be approximated by linear relations.
We use these specific altitudes for the calculation of the refractive index profile using  linear segmentation.
Such a procedure simplifies the calculation of atmospheric refraction but introduces  errors that grow for large zenith angles $Z\sim 90^\circ$.
The altitude dependence of the refractive index within the $i$th segment reads as
\begin{align}
\label{eq:refIndHeight}
n_i(h)=n_i+\left(\frac{\D n}{\D h}\right)_i(H_i-h).
\end{align}
Table~\ref{tab:RefIndex} summarizes the corresponding values of the refractive index values and gradients for the corresponding segments.

\begin{table}[h]
\caption{\label{tab:RefIndex}%
The defined reference levels, gradients, and values of the linearly segmented refractive index-height profiles from surface to 85 km.
}
\begin{ruledtabular}
\begin{tabular}{ l c c c }
\textrm{Sub-}&\textrm{Height}&
\textrm{Refractive index}&\textrm{Refractive index}\\
\textrm{script}&&
\textrm{gradient}&\\
$i$& $H_i$, (\textrm{km})&
$\left(\D n/\D h\right)_i{\times} 10^{{-}6}$, (\textrm{km}${}^{-1}$)& $(n_i-1)\times 10^8$\\
\colrule
0& 0 & -     & 27\,340 \\
1& 5 & 25.68 & 14\,660 \\
2& 7 & 17.58 & 11\,142 \\
3& 11& 12.50 & 6\,141  \\
4& 15& 7.183 & 3\,268  \\
5& 20& 3.565 & 1\,485  \\
6& 32& 1.042 & 235     \\
7& 47& 0.134 & 34      \\
8& 51& 0.034 & 21      \\
9& 71& 0.010 &  1      \\
10& 84.8& 0.001 & 0.1   \\
\end{tabular}
\end{ruledtabular}
\end{table}

For the calculation of the deterministic atmospheric extinction, the number density of absorbing and scattering particles is required.
The standard atmosphere model gives the following dependence of the relative number density on altitude $h$ (given in meters) \cite{Duntley1948},
\begin{align}
\label{eq:StAtmNdensity}
 N(h)/N_0=\exp\bigl[-h/\mathcal{H}_0\bigr].
\end{align}
Here $N_0$ is the number density at the observer level ($N_0=2.55\times 10^{25}$ m${}^{-3}$ at sea level) and $\mathcal{H}_0=6\,600$~m.

\section{Path elongation due to atmospheric refraction}
\label{app:refr}

\begin{figure}[ht]
 \includegraphics[width=0.45\textwidth]{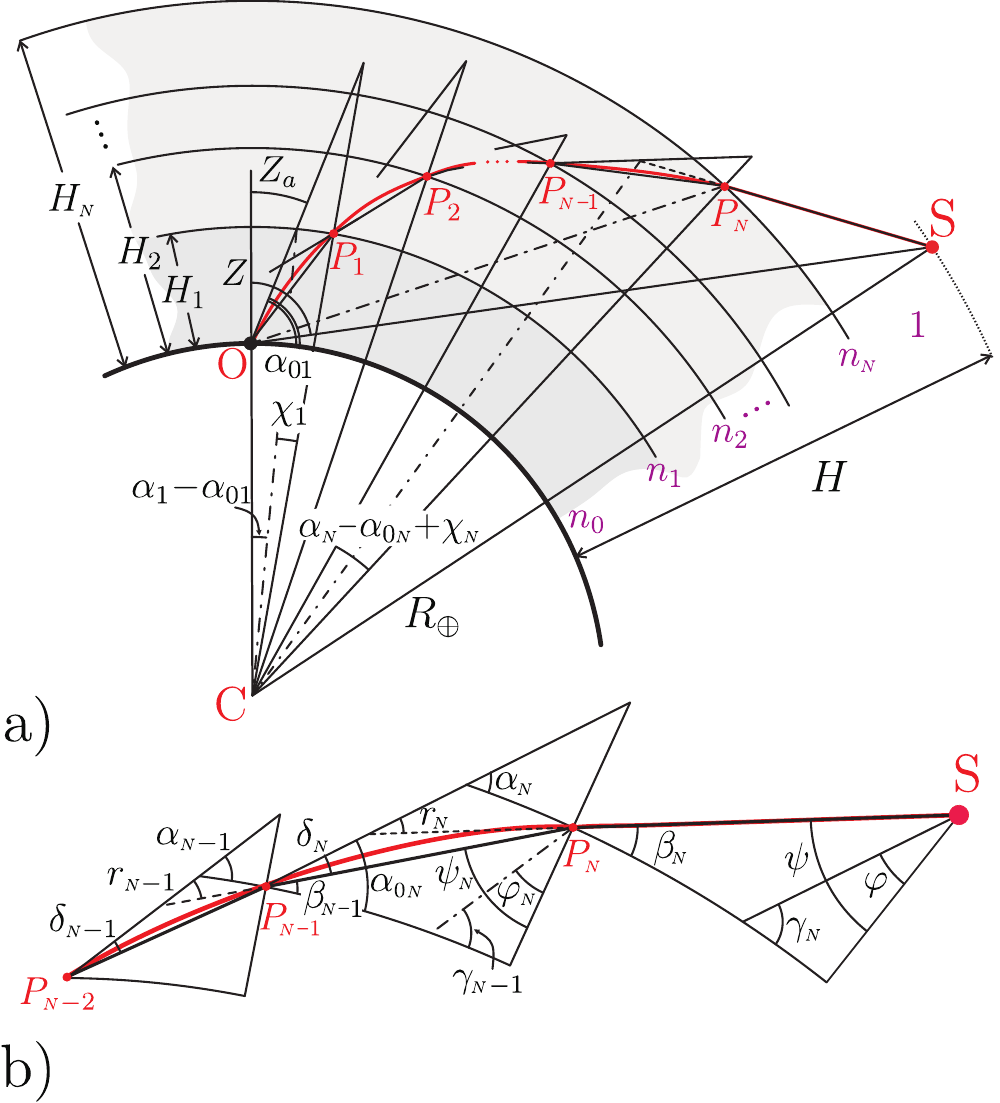}
 \caption{\label{fig:GRefraction} The influence of atmospheric refraction on the elongation of the optical ray trajectory (a). 
 The geometrical path of length $OS$ deforms into the curved path $OP_1P_2...P_nS$ while the true zenith angle $Z$ changes to the apparent zenith angle $Z_a$.
 The magnified upper part of the ray path shows the relevant refraction angles b).
}
\end{figure}

In the presence of Earth's atmosphere, the slant path $L$ is elongated due to refraction on interfaces of atmospheric layers with different refractive indices (see Fig.~\ref{fig:GRefraction}).
In the real atmosphere the gradient of the refractive index is a continuous function of height.
We use the standard atmosphere model and consider 10 atmospheric layers (c.f. Appendix~\ref{app:standardAtmosphere}, Table~\ref{tab:RefIndex}).
The refractive index is linearly segmented, such that it is a linear function of height  within one layer [cf. Eq.~(\ref{eq:refIndHeight})].

As a consequence of atmospheric refraction the apparent zenith angle $Z_a$ starts to deviate from the true  zenith angle $Z$.
These angles are related with each other as
\begin{align}
\label{eq:ZaZ}
 Z_a=\arcsin\left(\frac{1}{n_0}\sin Z\right),
\end{align}
where $n_0=1.00027$ is the refractive index of the lower layer of atmosphere~\cite{Birch1994}.
Near sea level the two zenith angles differ by approximately a minute of arc at $Z=45^\circ$ and  half a degree near the horizon.

Refraction of optical rays happens on the $i$th interface of two adjoined $i$th and $(i+1)$th layers of heights $H_i$ and $H_{i+1}$, correspondingly, and is characterized by the angle of incidence $\pi/2-\beta_i$ and the angle of refraction $\psi_i$ (see Fig.~\ref{fig:GRefraction}).
If the ray enters the observer telescope at the zenith angle $Z_a$, Snell's law yields the geometric invariant
\begin{align}
\label{eq:Snell}
 n_0 R_\oplus\cos\alpha_{01}=n_i(R_\oplus+H_i)\cos\beta_i=\text{const},
\end{align}
where
\begin{align}
 \alpha_{01}=\pi/2-Z_a
\end{align}
is the elevation angle at the observer $O$ and $R_\oplus$ is the Earth radius.
Using the notation
\begin{align}
 C_i=\frac{R_\oplus}{R_\oplus+H_i},\qquad C_H=\frac{R_\oplus}{R_\oplus+H}
\end{align}
with $H$ being the altitude of the satellite above the ground we derive from Eq.~(\ref{eq:Snell})
\begin{align}
 \beta_i=\arccos\left(\frac{n_0}{n_i}  C_i\sin Z_a\right)=\arccos\left( \frac{C_i}{n_i}\sin Z\right),\nonumber\\
  \beta_N=\arccos\left(n_0  C_H\sin Z_a\right)=\arccos\left( C_H\sin Z\right),
\end{align}
where the index $N$ corresponds to the last atmospheric layer at altitude 85 km which is accounted in our calculations (in our case $N=10$).
We  also use the index $i=1,...,N$ to denote the atmospheric layers above the ground level.

Using simple geometric considerations and Snell's law, we derive the following relations for the angles relevant for the calculation of the ray path length,
\begin{align}
 \alpha_i=\arccos\left(\frac{n_i}{n_{i-1}}\cos\beta_i\right),
\end{align}
\begin{align}
 \alpha_{0i}=\alpha_i-\alpha_{0(i-1)}+\beta_i+\chi_i+\psi_i-Z_a,\quad i\ne 1,
\end{align}
\begin{align}
 \chi_i=r_i-(\alpha_i-\beta_i),
\end{align}
\begin{align}
&\psi_1=\pi-Z_a-\delta_1-\alpha_1+\alpha_{01}-\chi_1,\\
& \psi_i=\arcsin\left(\frac{C_{i}}{C_{i-1}}\sin[Z_a-\beta_{i-1}+\alpha_{0(i-1)}]\right),\quad i\ne1,
\end{align}
\begin{align}
 \psi=\arcsin\left(\frac{C_H}{C_{N}}\sin[r_{N}-\delta_{N}+\psi_{N}]\right).
\end{align}
The remaining  angles to be determined are $\delta_i$ and $r_i$.
The former angle is associated with the local elevation angle error and  can be found from the law of sines
\begin{align}
 &(R_\oplus+H_{i-1})\cos\alpha_{0i}=(R_\oplus+H_i)\cos\alpha_i, \\
& (R_\oplus+H_{i-1})\cos\left(\alpha_{0i}-\delta_i\right)=(R_\oplus+H_i)\cos\left(\alpha_i+\chi_i-\delta_i\right).
\end{align}
Solving these equations with respect to $\delta_i$ we derive
\begin{align}
 \tan\delta_i&=\frac{\cos\alpha_i-\cos(\alpha_i+\chi_i)}{\sin(\alpha_i+\chi_i)-C_i/C_{i-1}\sin\alpha_{0i}},
\end{align}
\begin{align}
 \tan\delta_N&=\frac{\cos\alpha_N-\cos(\alpha_N+\chi_N)}{\sin(\alpha_N+\chi_N)-C_H/C_N\sin\alpha_{0N}}.
\end{align}
The remaining angle $r_i$ is the bending angle within an $i$th layer  which is calculated  from the refraction integral~\cite{Mahan1962},
\begin{align}
\label{eq:refr}
 r_i{=}\int\limits_{H_{i{-}1}}^{H_i}\D h\frac{1}{n}\frac{\D n}{\D h}
\frac{\cos\beta_{i-1}}{\sqrt{(n C_{i-1}/n_{i-1}C_i )^2{-}\cos^2\beta_{i-1}}}.
\end{align}
The  total angle $r=\sum_i r_i$ is known in optical astronomy as atmospheric refraction.
It can be shown that the refraction integral (\ref{eq:refr}) can be simplified to the following approximate analytical form,
\begin{align}
 r_i\approx2\frac{n_{i-1}-n_i}{\tan\beta_i+\tan\beta_{i-1}},
\end{align}
provided that $H_i-H_{i-1}\ll H_{i-1}$ and if the refractive index gradient $\D n/\D h$ is constant within the atmospheric layer.

The length of the optical ray path inside the $i$th layer can be found from the triangle $CP_{i-1}P_i$ applying the law of cosines
\begin{align}
\label{eq:LiMerid}
 &L_i{=}\Bigl\{(R_\oplus{+}H_{i{-}1})^2{+}(R_\oplus{+}H_i)^2\\
 &\qquad{-}2 (R_\oplus{+}H_{i-1})(R_\oplus{+}H_i)\cos\left[\Phi(Z_a, r_i)\right]\Bigr\}^{1/2},\nonumber
 \end{align}
 \begin{align}
 \label{eq:PhiApp}
 \Phi(Z_a, r_i)=\alpha_i{-}\alpha_{0i}{+}\chi_i.
\end{align}
For the optical path length in vacuum we find from the triangle $CP_NS$
\begin{align}
\label{eq:LN1Meridian}
 &L_{N+1}{=}\Bigl\{(R_\oplus{+}H_{N})^2{+}(R_\oplus{+}H)^2\\
 &\qquad{-}2 (R_\oplus{+}H_{N})
 (R_\oplus{+}H)\cos\left[ \Theta(Z_a,r_N)\right]\Bigr\}^{1/2}.\nonumber
\end{align}
Herein we need the angle
\begin{align}
\label{eq:ThetaApp}
 &\Theta(Z_a,r_N)=r_N{-}\delta_{N}{+}\psi_N-\psi.
\end{align}
The ratio of the total length of the ray trajectory to the geometric path length [c.f. Eq.~(\ref{eq:LcosZ})] reads as,
\begin{align}
\label{eq:elongation}
 \varepsilon_r=\frac{1}{L}\sum\limits_{i=1}^{N+1} L_i,
\end{align}
which is the elongation factor due to atmospheric refraction.

The elongation factor (\ref{eq:elongation}) is calculated without taking into account the finite radius of curvature of optical rays.
Actually, the beam curvature near the ground is the smallest one, i.e. the ray bending is the greatest.
In this case the empirical function for ray curvature $\mathcal{K}$ is given by~\cite{Gaifillia2016, Hirt2010}
 \begin{align}
 \mathcal{K}=R_\oplus\left\{670.87\frac{P}{T^2}\left[0.034+\lambda(h) 10^{-3}\right]\sin Z_a\right\}^{-1},
 \end{align}
where $T$  is the temperature in  K, $P$ is the pressure in mb, $\lambda(h)=\D T/\D h$ is the lapse rate in K/km.
For the standard atmosphere at $Z_a=90^\circ$ the curvature reaches its minimum value of $4.4 R_\oplus$.
Since this value is still larger than the Earth's radius, we can assume that the optical rays within each atmospheric layer can be considered to be straight lines.

\section{Atmospheric model of optical turbulence}
\label{app:Cn2models}

Dewan \textit{et al}. \cite{Dewan1993} proposed a simplified version of the multiparameter VanZandt model \cite{VanZandt1978} for the refractive index structure parameter variation with altitude.
This so-called Air Force Geophysics Laboratory (AFGL) model utilizes the meteorological data derived from radiosondes and yields the following analytic formula for the refractive-index structure parameter profile:
\begin{align}
\label{eq:AFGL}
 C_n^2(h)=2.8\left[M(h)\right]^2(0.1)^{\frac{4}{3}}10^{Y(h)}.
\end{align}
Here the factors $(0.1)^{4/3}10^{Y(h)}$ determine the outer scale of turbulence in a statistical manner.
The function $Y(h)$ is empirically related  with the altitude-dependent wind shear $S(h)$ and the lapse rate $\lambda(h)$ as \cite{Jackson2004}
\begin{widetext}
\begin{align}
Y(h)&=2.9767{+}27.9804 S(h){+}2.9012\lambda(h){+}1.1843\lambda(h)^2{+}0.1741\lambda(h)^3{+}0.0086\lambda(h)^4,\qquad (\text{lower troposphere})\\
&=0.7152{+}30.6024 S(h){+}0.0003\lambda(h){-}0.0057\lambda(h)^2{-}0.0016\lambda(h)^3{+}0.0001\lambda(h)^4,\qquad (\text{troposphere})\\
 &=0.6763{+}8.1569 S(h){-}0.0536\lambda(h){+}0.0084\lambda(h)^2{-}0.0007\lambda(h)^3{+}0.00002\lambda(h)^4. \qquad (\text{stratosphere}).
\end{align}
\end{widetext}
The parameter $M$ is connected with the gradient of the index of refraction,
\begin{align}
 M(h)=\frac{-79\times 10^{-6}P(h) N^2(h)}{g T(h)},
\end{align}
where $P$ is the pressure in mb, the temperature $T$ is given in K, the buoyancy frequency (Brunt-V\" ais\" al\" a frequency) in s$^{-1}$.
The buoyancy frequency  reads as
\begin{align}
 N^2=\frac{g\left[\lambda(h)+\gamma\right]}{T(h)},
\end{align}
where  $g$ is the acceleration of gravity, and $\gamma=9.8$ K/km is the dry air adiabatic lapse rate.
The altitude dependencies of pressure and temperature are governed by Eqs.~(\ref{eq:temper}), (\ref{eq:pressure}), and (\ref{eq:pressure1}).

\begin{figure}[ht]
 \includegraphics[width=0.45\textwidth]{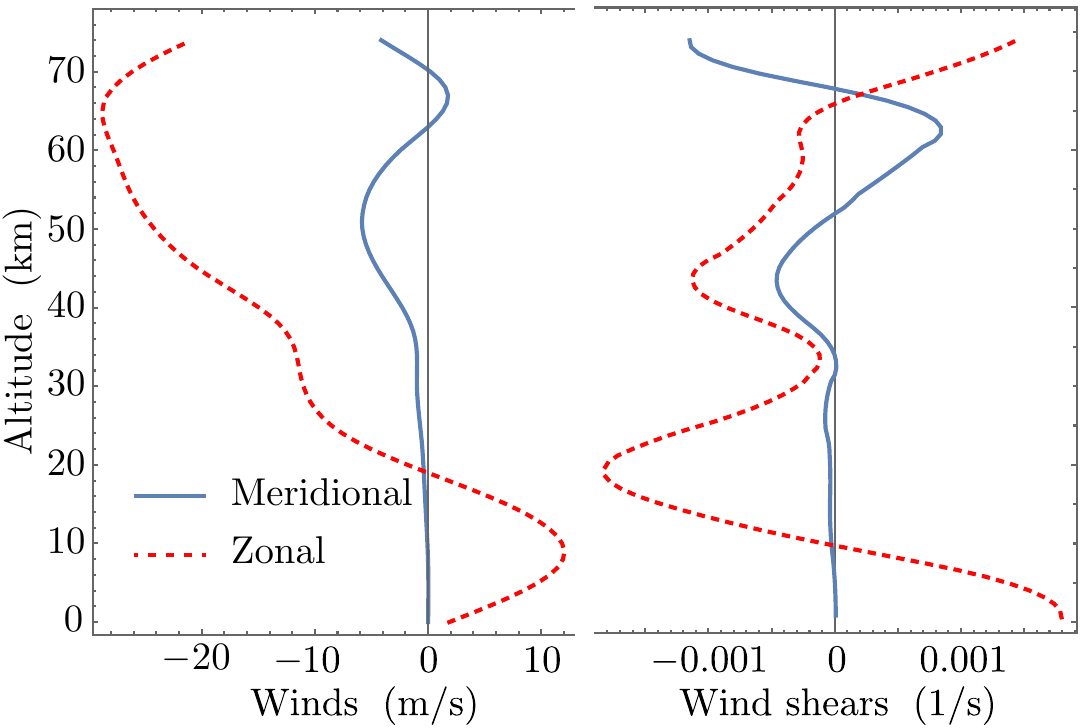}
 \caption{\label{fig:windNight} Typical example of the mean meridional (solid line) and  the zonal (dashed) winds is shown together with the corresponding vertical wind shear components.
The HWM93 empirical wind model~\cite{Hedin1996} for the northern hemisphere ($48^\circ$~N, $11.5^\circ$~E) summer (day 236) at midnight has been used for the calculation.
}
\end{figure}

The  altitude variation of the refractive index structure parameter can be determined from Eq.~(\ref{eq:AFGL}), provided the altitude variation of the wind shear $S(h)$ is known.
In this study we assume horizontal homogeneity of the atmosphere, which means that the mean wind properties do not depend on the horizontal position of the observer.
Thus, we assume a flat terrain and neglect any spatial inhomogeneity of Earth's surface.
This assumption is equivalent to the independence of the vertical wind component on altitude~\cite{Wyngaard}.
For the determination of the vertical wind shear components due to meridional and zonal winds we  use the HWM93 thermospheric wind model~\cite{Hedin1996}.
The examples of winds and corresponding shear components are shown in Fig.~\ref{fig:windNight} for summer night and the observer located near Munich~\cite{Program}.

For the atmospheric boundary layer close to the Earth's surface,  the refractive index structure parameter given by Eq.~(\ref{eq:AFGL}) is not applicable.
We use the Walters and Kunkel model (WK)~\cite{Walters1981} for the $C_n^2$ profile within the boundary layer,
\begin{align}
\label{eq:WKnight}
  \frac{C_n^2(h)}{C_n^2(h_0)}=(h/h_0)^{-2/3},  \qquad& h_0,h\le h_i
\end{align}
for nighttime and
\begin{align}
\label{eq:WKday}
 \frac{C_n^2(h)}{C_n^2(h_0)}=\begin{array}{l l}
                              (h/h_0)^{-4/3},                   \,& h_0,h\le0.5 h_i,\\
                              (0.5 h_i/h_0)^{-4/3}  ,                    \,& 0.5 h_i\le h\le 0.7 h_i,\\
                              2.9(0.5 h_i/h_0)^{-4/3}(h/h_i)^3, \,& 0.7 h_i\le h\le h_i
                             \end{array}
\end{align}
for daytime.
Here $h_i$ is the height of the inversion layer above the ground ($h_i\sim0.5$~km at nighttime and $\sim 1$~km at daytime), $h_0$ is the reference height referred to as the Monin-Obukhov scale ($h_0\sim 10$~m at nighttime and $\sim5$~m at daytime).

\begin{figure}[ht]
 \includegraphics[width=0.45\textwidth]{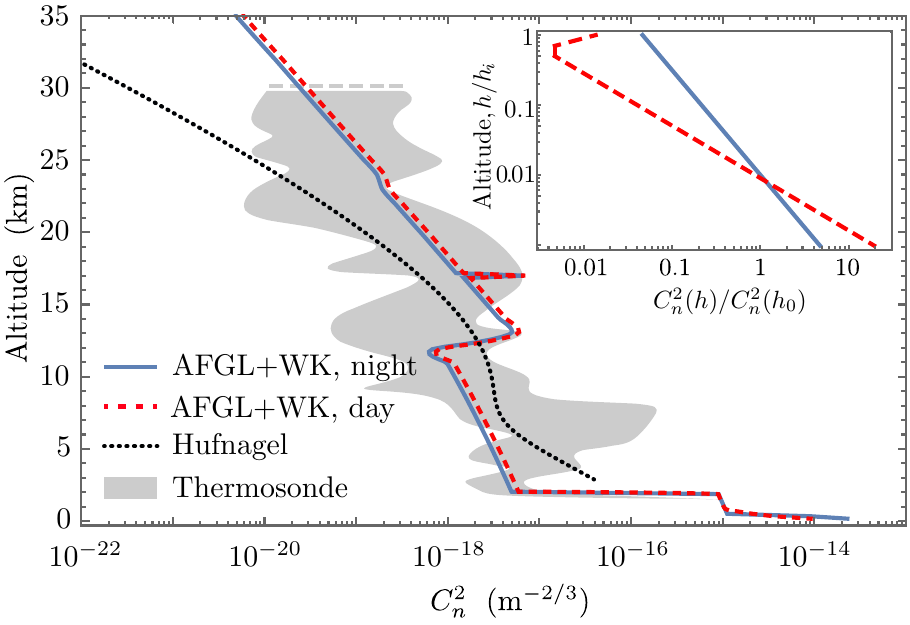}
 \caption{\label{fig:Cn2profile} Profiles of the refractive index structure function: AFGL, WK, and Hufnagel models are compared to the thermosonde data adopted from Ref.~\cite{Frehlich2010}.
 The inset shows the refractive-index structure parameter within the boundary layer with respect to the inversion height.
}
\end{figure}

Figure~\ref{fig:Cn2profile} shows the profile of the refractive-index structure parameter calculated using the AFGL and the WK models [cf. Eqs.~(\ref{eq:AFGL}), (\ref{eq:WKnight}), and (\ref{eq:WKday})].
For comparison, the widely used Hufnagel model~\cite{Hufnagel1964} as well as the estimation of $C_n^2$ based on thermosonde data~\cite{Frehlich2010} are shown.
The AFGL+WK model agrees better with the experimental profile than the Hufnagel model.
The difference between nighttime and daytime profiles is negligibly small for the altitudes above the inversion layer height $h_i$.
Figure~\ref{fig:windNight} shows the different behavior of the  wind shear for meridional and zonal winds.
However this anisotropy has almost no influence on the $C_n^2$ profile.
As a consequence, we could neglect the dependence of this turbulence characteristics on the direction of view of the observer.

Finally, we relate the vertical profile $C_n^2(h)$ to the corresponding profile along the slant range.
For the uplink configuration, the relation between the height $h$ of the certain point along the slant path and the distance between this point and the light source, $L_r\xi$, $\xi\in[0,1]$, can be found from the law of cosines [cf. Ref.~\cite{Gurvich} and see Fig.~\ref{fig:Geometry1} (a)]
\begin{align}
h^{\mathrm{UL}}(\xi)&=R_\oplus\sqrt{1+2\frac{L\xi}{R_\oplus}\cos Z+\frac{(L\xi)^2}{R_\oplus^2}}-R_\oplus\\
&\approx R_\oplus\sqrt{1+2\frac{L_r\xi}{R_\oplus}\cos Z_a+\frac{(L_r\xi)^2}{R_\oplus^2}}-R_\oplus.\nonumber
\end{align}
Similarly, for the downlink we derive by replacing $\xi\rightarrow1-\xi$
\begin{align}
h^{\mathrm{DL}}(\xi)&\approx R_\oplus\sqrt{1+2\frac{L_r[1-\xi]}{R_\oplus}\cos Z_a+\frac{(L_r[1-\xi])^2}{R_\oplus^2}}\nonumber\\
&-R_\oplus.
\end{align}
Under the condition $L_r/R_\oplus\ll 1$, these equations reduce to
\begin{align}
\label{eq:hUL}
h^{\mathrm{UL}}(\xi)\approx L_r\xi\cos Z_a,
\end{align}
\begin{align}
\label{eq:hDL}
h^{\mathrm{DL}}(\xi)\approx L_r(1-\xi)\cos Z_a.
\end{align}
The refractive index structure parameter given by Eqs.~(\ref{eq:AFGL}), (\ref{eq:WKnight}), and (\ref{eq:WKday}) maps for the uplink slant range as $C_n^2(h)\rightarrow C_n^2(h^{\mathrm{UL}})= C_n^2(L_r,\xi)$ and for the downlink as $C_n^2(h)\rightarrow  C_n^2(L_r,1-\xi)$.

\section{Approximation of Eq.~(\ref{eq:Jkernel}) for satellite-mediated links}
\label{app:Kernel}

In this appendix we give the approximation for the integral kernel (\ref{eq:Jkernel}).
We consider the downlink configuration and just note that the formulas for uplink are obtained in the similar footing with the variable replacement $\xi\rightarrow 1-\xi$.
For the case of strong turbulence or long propagation distances in turbulence the coherence radius $\rho_0$ [cf.~Eq.~(\ref{eq:rho0})] is small.
This condition is well satisfied for satellite-mediated atmospheric links.
In this case, the exponential in Eq.~(\ref{eq:Jkernel}), $\mathcal{J}(\mathbf{r},\mathbf{r}_1^\prime,\mathbf{r}_2^\prime,\mathbf{r}_3^\prime)$, differs significantly from zero in the following regions:
\begin{align}
\label{region1}
 |\mathbf{r}_2^\prime|\xi\gg\rho_0,\quad |\mathbf{r}_3^\prime|\xi,\,\,|\mathbf{r}(1{-}\xi){+}\mathbf{r}_1^\prime\xi|\lesssim\rho_0;
\end{align}
\begin{align}
\label{region2}
 |\mathbf{r}(1{-}\xi){+}\mathbf{r}_1^\prime \xi |\gg\rho_0,\quad|\mathbf{r}_2^\prime|\xi,\,|\mathbf{r}_3^\prime|\xi\lesssim\rho_0;
 \end{align}
 \begin{align}
 \label{region3}
 |\mathbf{r}_2^\prime|\xi,\,\, |\mathbf{r}_3^\prime|\xi,\,\,|\mathbf{r}(1{-}\xi){+}\mathbf{r}_1^\prime\xi|\lesssim\rho_0.
\end{align}
The function  (\ref{eq:Jkernel}) can be approximated then as linear combination of three terms~\cite{Banakh1979, Charnotskii2010}
\begin{align}
 \mathcal{J}=\mathcal{J}_1+\mathcal{J}_2-\mathcal{J}_3,
\end{align}
\begin{widetext}
\begin{align}
\label{eq:1term}
 &\mathcal{J}_1(\mathbf{r},\mathbf{r}_1^\prime,\mathbf{r}_2^\prime,\mathbf{r}_3^\prime)=\exp\Bigl[-\rho_0^{-\frac{5}{3}}\int\limits_0^1\!\!\D\xi\frac{C_n^2( L_r,1{-}\xi)}{C_{n,0}^2}\sum\limits_{j=1,2}\left|\mathbf{r}(1{-}\xi){+}[\mathbf{r}_1^\prime{+}(-1)^j\mathbf{r}_3^\prime]\xi\right|^{\frac{5}{3}}\Bigr]\\
&\times
\sum\limits_{n=0}^\infty\frac{\rho_0^{-\frac{5}{3}n}}{n!}\Bigl\{\int\limits_0^1\!\D\xi \frac{C_n^2(L_r,1{-}\xi)}{C_{n,0}^2}\sum\limits_{j=1,2}\Bigl(
\left|\mathbf{r}(1{-}\xi){+}[\mathbf{r}_1^\prime{+}(-1)^j\mathbf{r}_2^\prime]\xi\right|^{\frac{5}{3}}- \left|[\mathbf{r}_2^\prime{+}(-1)^j\mathbf{r}_3^\prime]\xi\right|^{\frac{5}{3}}\Bigr)\Bigr\}^n,
\nonumber
\end{align}
\begin{align}
\label{eq:2term}
 &\mathcal{J}_2(\mathbf{r},\mathbf{r}_1^\prime,\mathbf{r}_2^\prime,\mathbf{r}_3^\prime)=\exp\Bigl[-\rho_0^{-\frac{5}{3}}\int\limits_0^1\!\!\D\xi\frac{C_n^2( L_r,1{-}\xi)}{C_{n,0}^2}\left|[\mathbf{r}_2^\prime{+}(-1)^j\mathbf{r}_3^\prime]\xi\right|^{\frac{5}{3}}\Bigr]\\
 &\times\sum\limits_{n=0}^\infty\frac{\rho_0^{-\frac{5}{3}n}}{n!}\Bigl\{\int\limits_0^1\D\xi \frac{C_n^2( L_r,1{-}\xi)}{C_{n,0}^2}\sum\limits_{j=1,2}\Bigl(\left|\mathbf{r}(1{-}\xi){+}[\mathbf{r}_1^\prime{+}(-1)^j\mathbf{r}_2^\prime]\xi\right|^{\frac{5}{3}}-\left|\mathbf{r}(1{-}\xi){+}[\mathbf{r}_1^\prime{+}(-1)^j\mathbf{r}_3^\prime]\xi\right|^{\frac{5}{3}}\Bigr)\Bigr\}^n,\nonumber
\end{align}
\begin{align}
\label{eq:3term}
  &\mathcal{J}_3(\mathbf{r},\mathbf{r}_1^\prime,\mathbf{r}_2^\prime,\mathbf{r}_3^\prime)=\exp\Bigl[-\rho_0^{-\frac{5}{3}}\int\limits_0^1\D\xi \frac{C_n^2(L_r,1{-}\xi)}{C_{n,0}^2}\sum\limits_{j=1,2}\Bigl\{ \left|[\mathbf{r}_2^\prime{+}(-1)^j\mathbf{r}_3^\prime]\xi\right|^{\frac{5}{3}}\\
&+\left|\mathbf{r}(1{-}\xi){+}[\mathbf{r}_1^\prime{+}(-1)^j\mathbf{r}_3^\prime]\xi\right|^{\frac{5}{3}}\Bigr\}\Bigr]\sum\limits_{n=0}^\infty\frac{\rho_0^{-\frac{5}{3}n}}{n!}\Bigl\{\int\limits_0^1\D\xi\frac{C_n^2(
L_r,1{-}\xi)}{C_{n,0}^2}\sum\limits_{j=1,2}\left|\mathbf{r}(1{-}\xi){+}[\mathbf{r}_1^\prime{+}(-1)^j\mathbf{r}_2^\prime]\xi\right|^{\frac{5}{3}}\Bigr\}^n.\nonumber
\end{align}
The  first term (\ref{eq:1term}) accounts for the contributions from regions (\ref{region1}) and (\ref{region3}).
The term (\ref{eq:2term}) accounts for the regions (\ref{region2}) and (\ref{region3}), while the term (\ref{eq:3term}) eliminates the double counting of the region (\ref{region3}) in the integral kernel $\mathcal{J}$.
In the first approximation  ($n=0$) we have
\begin{align}\label{expans1}
&\mathcal{J}(\mathbf{r},\mathbf{r}_1^\prime,\mathbf{r}_2^\prime,\mathbf{r}_3^\prime)\nonumber\\
&\approx\exp\Bigl[-\rho_0^{-\frac{5}{3}}\int\limits_0^1\!\!\D\xi\frac{C_n^2( L_r,1{-}\xi)}{C_{n,0}^2}\sum\limits_{j=1,2}\left|\mathbf{r}(1{-}\xi){+}[\mathbf{r}_1^\prime{+}(-1)^j\mathbf{r}_3^\prime]\xi\right|^{\frac{5}{3}}\Bigr]+\exp\Bigl[-\rho_0^{-\frac{5}{3}}\int\limits_0^1\!\!\D\xi\frac{C_n^2(
L_r, 1{-}\xi)}{C_{n,0}^2}\left|[\mathbf{r}_2^\prime{+}(-1)^j\mathbf{r}_3^\prime]\xi\right|^{\frac{5}{3}}\Bigr]\nonumber\\
&\quad-\exp\Bigl[-\rho_0^{-\frac{5}{3}}\int\limits_0^1\D\xi \frac{C_n^2( L_r,1{-}\xi))}{C_{n,0}^2}\sum\limits_{j=1,2}\Bigl\{ \left|[\mathbf{r}_2^\prime{+}(-1)^j\mathbf{r}_3^\prime]\xi\right|^{\frac{5}{3}}+\left|\mathbf{r}(1-\xi){+}[\mathbf{r}_1^\prime{+}(-1)^j\mathbf{r}_3^\prime]\xi\right|^{\frac{5}{3}}\Bigr\}\Bigr],
\end{align}
which already gives a good approximation to the kernel $\mathcal{J}$ (cf. Ref.~\cite{Banakh1979}).
\end{widetext}

\section{Phenomenological model of aperture-averaged scintillations}
\label{app:Young}

The rigorous analysis of satellite-mediated quantum communication links should account for turbulent disturbances for a wide range of the zenith angle.
Here we calculate the aperture averaged scintillation index with  accounting for saturation effects.
The aperture-averaged scintillation index can be obtained by substitution of Eq.~(\ref{eq:eta}) in Eq.~(\ref{eq:scintEta}), which leads to~\cite{Tatarskii2016}
\footnote{Originally, the expression for the scintillation index for the log transmittance, $\sigma_{\log\eta}^2$, is derived in the  chapter 13 of Ref.~\cite{Tatarskii2016}.
Equation (\ref{eq:sciIndex}) is then obtained from the relation $\sigma_\eta^2=\exp[\sigma_{\log\eta}^2]-1$ (cf. Ref.~\cite{Andrews2005}). }
\begin{align}
\label{eq:sciIndex}
 \sigma^2_\eta&=\frac{1}{\langle I\rangle^2}\int_{\mathbb{R}^2} \D^2\boldsymbol{\kappa} F_I(\boldsymbol{\kappa},L_r) f_\mathcal{A}(\boldsymbol{\kappa}),
\end{align}
where
\begin{align}
\label{eq:ApertureFilter}
 f_{\mathcal{A}}(\boldsymbol{\kappa})=\left|\frac{1}{\mathcal{A}}\int_{\mathcal{A}}\D^2\boldsymbol{r} e^{i\boldsymbol{\kappa}\cdot\boldsymbol{r}}\right|^2
\end{align}
is the aperture filter function, and $\langle I\rangle$ is the mean intensity.
The Fourier transformed correlation function of the intensity fluctuations is given by
\begin{align}
\label{eq:intCorr}
 &F_I(\boldsymbol{\kappa},L_r)=\int\limits_{\mathbb{R}^2}\D^2\boldsymbol{r} B_I(\boldsymbol{r},L_r) e^{-i\boldsymbol{\kappa}\cdot\boldsymbol{r}},\\
 &\quad B_I(\boldsymbol{r}_1{-}\boldsymbol{r}_2,L_r) =\langle \left[I(\boldsymbol{r}_1,L_r){-}\langle I\rangle\right]\left[I(\boldsymbol{r}_2,L_r){-}\langle I\rangle\right]\rangle.\nonumber
\end{align}
For a circular aperture of radius $a$ the filter function~(\ref{eq:ApertureFilter}) is easily evaluated to be
\begin{align}
\label{eq:ApertureFilterCircular}
 f_{\mathcal{A}}(\boldsymbol{\kappa})=f_{a}(\boldsymbol{\kappa})=\left[\frac{2 J_1(\kappa a)}{\kappa a}\right]^2,
\end{align}
where $\kappa=|\boldsymbol{\kappa}|$ and $J_n(x)$ is a Bessel function of the first kind.

In the limiting case of the vanishingly small Fresnel number $\Omega=kW_0^2/(2L_r)\rightarrow 0$ the collimated beam with the initial beam spot size $W_0$ can be considered as a plane wave.
Indeed, for the satellite-mediated link under consideration the slant range $L_r$ is large enough even at the zenith and $\Omega\rightarrow 0$.
In this limiting case the intensity spectral density $F_I(\boldsymbol{\kappa},L_r)$ can be  obtained by solving the equations of geometrical optics~\cite{Tatarskii2016}\footnote{Reference~\cite{Tatarskii2016} derives the formula for the logarithmic amplitude of the light wave $\chi=\log(A/A_0)$.
Under weak fluctuation conditions the relation $F_I(\boldsymbol{\kappa},L_r)=4\langle I\rangle^2F_\chi(\boldsymbol{\kappa},L_r)$ is valid~\cite{Andrews2005}. }
\begin{align}
\label{eq:FIdivergent}
 F_I(\boldsymbol{\kappa},L_r)&=2\pi\langle I\rangle^2\kappa^4\int\limits_0^{L_r}\D z {z}^2\Phi_n(\kappa,z),
\end{align}
where the  turbulence spectrum $\Phi_n$ is given in Eq.~(\ref{eq:PhiTurb}).
For the Kolmogorov turbulence spectrum (\ref{eq:Kolmogorov}) the intensity spectral density $ F_I(\boldsymbol{\kappa},L_r)$ is proportional to $\kappa^{1/3}$ in the inertial range.
Thus, for large values of the transverse wave number the intensity spectral function diverges.
In order to remedy this nonphysical effect we introduce the cutoff of higher spatial frequencies $\Delta\kappa$ such that $\kappa\in[0,\Delta\kappa]$.
The specifical choice of $\Delta\kappa$ will be discussed later in this appendix.
Taking into account this cutoff effect the aperture averaged scintillation index is calculated  straightforwardly,
\begin{align}
\label{eq:sigmaEta1}
  \sigma^2_\eta=\frac{16\pi^2 }{a^2}\int\limits_0^{L_r}\D z {z}^2\int\limits_{0}^{\Delta\kappa}\D \kappa \kappa^{3}\Phi_n(\kappa,z)\left[J_1(\kappa a) \right]^2.
\end{align}
The obtained result shows that the intensity fluctuations are directly related to spectrum of turbulent fluctuations of the refractive index.

For the Kolmogorov turbulence spectrum (\ref{eq:Kolmogorov}) the integrations can be performed, provided the altitude dependence of structure constant is known.
Here we adopt a simple dependence on the $z$ variable  assuming that the strongest turbulence is near the Earth surface and decreases exponentially with the height~\cite{Young1969},
\begin{align}
\label{eq:CnH0}
 C_n^2(z)=C_{n,0}^2\exp\left[-\frac{z}{H_0\sec Z_a}\right].
\end{align}
Here $H_0$ is the characteristic length at zenith and the factor $\sec Z_a$ accounts for the increase of this length for the slant paths.
The integral with respect to  $z$ in Eq.~(\ref{eq:sigmaEta1}) is evaluated as
\begin{align}
\label{eq:mathcalI}
 \mathcal{I}&(Z_a)=2(H_0\sec Z_a)^3-\exp\left[-\frac{L_r(Z_a)}{H_0\sec Z_a}\right]H_0\sec Z_a\nonumber\\
 &{\times}\left\{2(H_0\sec Z_a)^2+2 H_0\sec Z_a L_r(Z_a)+L_r^2(Z_a)\right\}.
\end{align}
In the limit of infinitely distant light source, $L_r\rightarrow\infty$, we get $ \mathcal{I}(Z_a)=2(H_0\sec Z_a)^3$.
Consequently, the scintillation index (\ref{eq:sigmaEta1}) for $\Delta\kappa\rightarrow\infty$ has the same dependence on the zenith angle as those obtained within the Rytov approximation, cf. Eq.~(\ref{eq:scintEta}).

The cutoff spatial frequency $\Delta\kappa$ in Eq.~(\ref{eq:sigmaEta1}) can be estimated from the following considerations.
The rigorous asymptotic analysis shows~\cite{Fante1983} that for long propagation paths the covariance function $B_I(\boldsymbol{r},L_r)$ in Eq.~(\ref{eq:FIdivergent}) differs from zero if $|\boldsymbol{r}|\le \Delta r$, where
\begin{align}
\label{eq:deltar}
\Delta r=0.18 \lambda L_{\mathrm{turb}}/\rho_0.
\end{align}
Here, $\lambda$ is the optical wavelength, $L_{\mathrm{turb}}$ is the propagation path length in turbulence, and $\rho_0$ is the spatial coherence length given by Eq.~(\ref{eq:rho0}).
According to Eq.~(\ref{eq:CnH0}) the turbulence is sufficient for heights $h\le H_0$.
Hence, without loss of generality we set $L_{\mathrm{turb}}=H_0\sec Z_a$.
The finiteness of the definition domain for the covariance function $B_I(\boldsymbol{r},L_r)$ introduces the cutoff frequency $\Delta\kappa$ for the spectral density $F_I(\boldsymbol{\kappa},L_r)$.
Using the relation of Fourier analysis $\Delta\kappa\Delta r\ge\mu$, we may estimate $\Delta\kappa=\mu/\Delta r$, where $\mu\sim 1$ is a phenomenological parameter.
It is easy to see that $\Delta\kappa\propto (\sec Z_a)^{-8/5}$ and decreases while the zenith angle approaches the horizon.
For further insight into physical meaning of the parameters $\Delta r$ and $\Delta\kappa$ we point the reader's attention on Ref.~\cite{Andrews1999}.
Finally, performing the integration of (\ref{eq:sigmaEta1}) over $\kappa$ and over $z$ variable [cf. Eq.~(\ref{eq:mathcalI})] we obtain Eq.~(\ref{eq:sigmaEta2text}),
where only the first term of $\mathcal{I}(Z_a)$ is accounted.
At the cost of analytic simplicity, the obtained result can be further generalized for the case of  Cassegrain-type apertures with the inner circular obscuration \cite{Vasylyev2018a}.

\section{Decoy state: statistical fluctuation analysis}
\label{app:Decoy}

In this appendix, we summarize the method for estimation the lower (upper) bounds of single-photon gain  and error rate taking into account the finite-key effects~\cite{Zhang2017}.
On quantum state preparation stage, Alice generates each bit in her raw key by randomly choosing the encoding basis ($X$ or $Z$) and the intensity (corresponding to the vacuum state, weak decoy state, and signal state).
The total number of bits sent by Alice to Bob is given by
\begin{align}
 N=N^s+N^d+N^v,
\end{align}
where the superscripts correspond to signal ($s$), weak-decoy ($d$), and vacuum ($v$) states.
We denote   $q^a=N^a/N$ as the rate with which Alice encodes a state with intensity $\mu_a$, $a=s,d,v$.
The conditional probability that an $i$-photon state corresponds to a coherent pulse with the intensity $\mu_a$ is
\begin{align}
\label{eq:condProbPia}
 p_i^a\approx\frac{N_i^a}{N_i}=\frac{N^ae^{-\mu_a}(\mu_a)^i/i!}{\sum_{\alpha\in\{s,d,v\}}N^\alpha e^{-\mu_\alpha}(\mu_\alpha)^i/i!},
\end{align}
where the approximation sign is due to statistical fluctuations.
The finiteness of numbers of generated bits yields
\begin{align}
 q=\frac{N^s}{2N}
\end{align}
for the $q$ parameter in Eq.~(\ref{eq:keyrate}).

Bob measures the received states in $X$ or $Z$ basis chosen randomly.
After basis reconciliation and key sifting, Bob possesses the total number of sifted bits
\begin{align}
 M=M^s+M^d+M^v.
\end{align}
For the atmospheric quantum channel, the bit numbers $N$ and $M$ are related as
\begin{align}
M = \frac{1}{2}\eta_{\mathrm{d}} \eta N,
\end{align}
where the factor $1/2$ is due to the sifting procedure, $\eta_{\mathrm{d}}$ is the deterministic channel loss including the detector efficiency given by Eq.~(\ref{eq:etad}), and $\eta$ is the fluctuating channel transmittance.
The bit number corresponding to the $i$-photon state is determined as
\begin{align}
M_i^a\approx p_i^a M_i = p_i^a \sum_{\alpha\in\{s,d,v\}}M^\alpha e^{-\mu_\alpha}(\mu_\alpha)^i/i!,
\end{align}
where $p_1^a$ is the same as the probability (\ref{eq:condProbPia}) chosen by Alice.

After error correction and error verification the secure key rate depends on a lower bound, on the gain of single-photon components of the signal state $Q_1^{L}$ and an upper bound, on the corresponding error rate $e_1^{U}$.
The estimation of $Q_1^{L}$ and $e_1^{U}$ should be performed in each basis.
The corresponding components we denote by superscripts $x$ and $z$ referring to the $X$ and $Z$ bases, correspondingly.
For the $Z$ basis the bounds on the single-photon gain and the error rate are given by
\begin{align}
 Q_1^{zL} =  Y_1^{zL} \mu_s e^{-\mu_s},
\end{align}
\begin{align}
\label{eq:e1zU}
 e_1^{zU}=e_1^{xU}+\theta^U,
\end{align}
where
\begin{align}
\label{eq:yield}
 Y_1^{\gamma L}=\frac{\mu_s}{\mu_s\mu_d-\mu_d^2}\Bigl(Q_{\mu_d}^{\gamma L}e^{\mu_d}&-\frac{\mu_d^2}{\mu_s^2}Q_{\mu_s}^{\gamma U}e^{\mu_s}\\
 &-\frac{\mu_s^2-\mu_d^2}{\mu_s^2}Y_0^U\Bigr),\quad\gamma=x,z,\nonumber
\end{align}
\begin{align}
\label{eq:errorxU}
 e_1^{xU}=\frac{(E_{\mu_d}Q_{\mu_d}^x)^Ue^{\mu_d}-e_0Y_0^L}{\mu_d Y_1^{xL}}.
\end{align}
Here $e_0=1/2$ and the  lower and upper bounds
\begin{align}
 Q_{\mu_d}^{zL} = \frac{Q_{\mu_d}^z}{1+\delta(M^{dz})} = \frac{Q_{\mu_d}^z}{1+\delta(N^{dz}Q_{\mu_d}^z)},
\end{align}
\begin{align}
 Q_{\mu_s}^{zU} = \frac{Q_{\mu_s}^z}{1-\delta(N^{sz}Q_{\mu_s}^z)},
\end{align}
\begin{align}
  Y_0^{L} = \frac{Y_0}{1+\delta(N^v Y_0)},\quad Y_0^{U} = \frac{Y_0}{1-\delta(N^v Y_0)},
\end{align}
\begin{align}
 (E_{\mu_d}Q_{\mu_d}^x)^U=\frac{E_{\mu_d}^xQ_{\mu_d}^x}{1-\delta(N^{dx} E_{\mu_d}^xQ_{\mu_d}^x)}
\end{align}
 are estimated by using the Chernoff bound method and are related to the overall gain (\ref{eq:OverallGain}), with the overall quantum bit error rate (\ref{eq:OverallQBER}) components, and with
  the counting rate for vacuum  decoy states and dark count contributions
 \begin{align}
  Y_0=\frac{M^v}{N(e^{-\mu_s}q^s+e^{-\mu_d}q^d+q^v)}+Y_0^{\mathrm{DC}}.
 \end{align}
 The deviation function
\begin{align}
 \delta(x)=\frac{-3\ln(\varepsilon/2)+\sqrt{[\ln(\varepsilon/2)]^2-8\ln(\varepsilon/2)x}}{2[x+\log(\varepsilon/2)]}
\end{align}
can be determined for the specified failure probability $\varepsilon$.

Finally, the upper bound $\theta^U$ in (\ref{eq:e1zU}) is obtained  by numerically solving
\begin{align}
\label{eq:epsilonTheta}
 \varepsilon&=\frac{\sqrt{\langle M_1^{xL}\rangle_{\mathrm{tr}}+\langle M_1^{szL}\rangle_\mathrm{tr}}}{\sqrt{\langle e_1^{xU}\rangle_\mathrm{tr}(1-\langle e_1^{xU}\rangle_\mathrm{tr})\langle M_1^{xL}\rangle_\mathrm{tr}\langle M_1^{szL}\rangle_\mathrm{tr}}}\\
 &\qquad\times 2^{-(\langle M_1^{xL}\rangle_\mathrm{tr}+\langle M_1^{szL}\rangle_\mathrm{tr})\xi(\theta^U)}\nonumber
\end{align}
with respect to $\theta^U$ ~\cite{Fung2010, Zhang2017}.
Here, the averaging $\langle ... \rangle_\mathrm{tr}$ is performed according to Eq.~(\ref{eq:etatr}) and the lower bounds of sifted key numbers are
\begin{align}
\label{eq:M1L}
 M_1^{\gamma L}=Y_1^{\gamma L}N(e^{-\mu_s}\mu_s q^s+e^{-\mu_d}\mu_d q^d),\quad\gamma=x,z,
\end{align}
\begin{align}
 M_1^{szL}=[1-\delta(p_1^s M_1^{zL})]p_1^sM_1^{zL}
\end{align}
with $q^a=N^a/N$.
In Eq.~(\ref{eq:epsilonTheta}) the exponential function $\xi(\theta)$ read as \cite{Fung2010}
\begin{align}
 \xi(\theta)\approx \frac{\ln 2}{2}\frac{(1-q^x)q^x}{(1-\langle  e_1^{xU}\rangle_\mathrm{tr})\langle  e_1^{xU}\rangle_\mathrm{tr}}\theta^2,
\end{align}
where $q^x{=}\langle M_1^{xL}\rangle_\mathrm{tr}/(\langle M_1^{xL}\rangle_\mathrm{tr}{+}\langle M_1^{szL}\rangle_\mathrm{tr})$ is the bias ratio.

\end{document}